\documentclass[aps,pre,twocolumn,superscriptaddress,showpacs]{revtex4-1}

\usepackage[english]{babel}
\usepackage[utf8x]{inputenc}
\usepackage{amsmath,amsfonts,amssymb,amsbsy,eucal,dsfont}
\usepackage{natbib}
\usepackage{color}
\usepackage[colorlinks]{hyperref}
\usepackage{graphicx}
\usepackage{rotating}
\newcommand{\hide}[1]{}

\begin{document}

\title{Highly asymmetric electrolytes in the primitive model:\\
Hypernetted chain solution in arbitrary spatial dimensions}
%\title{Hypernetted chain solution of the highly asymmetric primitive electrolyte model in arbitrary spatial dimensions}

\author{Marco Heinen}
\email[]{mheinen@thphy.uni-duesseldorf.de}
\affiliation{Institut f\"{u}r Theoretische Physik II, Weiche Materie, Heinrich-Heine-Universit\"{a}t D\"{u}sseldorf, 40225 D\"{u}sseldorf, Germany}

\author{Elshad Allahyarov}
\affiliation{Institut f\"{u}r Theoretische Physik II, Weiche Materie, Heinrich-Heine-Universit\"{a}t D\"{u}sseldorf, 40225 D\"{u}sseldorf, Germany}
\affiliation{Theoretical Department, Joint Institute for High Temperatures, Russian Academy of Sciences (IVTAN), 13/19 Izhorskaya street, Moscow 125412, Russia}

\author{Hartmut L\"owen}
\affiliation{Institut f\"{u}r Theoretische Physik II, Weiche Materie, Heinrich-Heine-Universit\"{a}t D\"{u}sseldorf, 40225 D\"{u}sseldorf, Germany}

\date{\today}

%%%%%%%%%%%%%%%%%%%%%%%%%%%%%%%%%%%%%%%%%%%%%%%%%%%%%%%%%%%%%%%%%%%%%%%%%
\renewcommand{\figurename}{Fig. }
\renewcommand{\figuresname}{Figs. }
\renewcommand{\refname}{Ref. }
\newcommand{\refsname}{Refs. }
\newcommand{\expressionname}{Eq. }
\newcommand{\expressionsname}{Eqs. }
\newcommand{\sectionname}{Sec. }
\newcommand{\sectionsname}{Secs. }
%%%%%%%%%%%%%%%%%%%%%%%%%%%%%%%%%%%%%%%%%%%%%%%%%%%%%%%%%%%%%%%%%%%%%%%%%

\begin{abstract}
The pair-correlation functions for fluid ionic mixtures in arbitrary spatial dimensions are
computed in hypernetted chain (HNC) approximation.
In the primitive model, all ions are approximated as non-overlapping hyperspheres with Coulomb interactions.
Our spectral HNC solver is based on a Fourier-Bessel transform introduced by Talman [J. Comput. Phys., \textbf{29}, 35 (1978)],
with logarithmically spaced computational grids.
Numeric efficiency for arbitrary spatial dimensions is a commonly exploited virtue of this transform method.
Here, we highlight another advantage of logarithmic grids, consisting in efficient sampling
of pair-correlation functions for highly asymmetric ionic mixtures.
For three-dimensional fluids, ion size- and charge-ratios larger than one thousand can be treated,
corresponding to hitherto computationally not accessed micrometer-sized colloidal spheres in 1-1 electrolyte.
Effective colloidal charge numbers are extracted from our primitive model results.
For moderately large ion size- and charge-asymmetries, we present Molecular Dynamics simulation results
that agree well with the approximate HNC pair correlations.
\end{abstract}

%\pacs{82.70.Kj, %Emulsions and suspensions
%      82.70.Dd, %Colloids
%      }

\maketitle

\section{Introduction}\label{sec:Intro}
%%%%%%%%%%%%%%%%%%%%%%%%%%%%%%%%%%%%%%%%%%%%%%%%%%%%%%%%%%%%%%%%%%%%%%%%%%%%%%%%%%%%%%%%%%%%%%%%%%%%%%%%%%%%%%%%
Most of the essential features of polyelectrolyte solutions can be efficiently modeled by a combination of charges and excluded volume, see 
\cite{Levin2002, Hansen2000, Messina2009, Vlachy1999} for reviews.
In the so-called primitive model (PM), all specific properties of the solvent are neglected except for its dielectric constant.
Research over the past decades has shown that some of the basic properties of polyelectrolytes (like screening and Coulomb association)
are contained in the asymmetric PM of electrolytes which enables wide applications to
charged colloidal suspensions, micelles and globular proteins.
However, for high asymmetry in charge and size between the microions and macroions, as occurring in suspensions of charged colloidal particles,
the PM is not easy to solve numerically in general.
For example, structural correlations in the PM were obtained by numerically expensive computer simulations only up to charge and diameter
asymmetries of about 1:100 \cite{Lobaskin1999, Linse1999, Hynninen2005, Allahyarov2009}, corresponding to the micellar rather than the colloidal regime.

In the present paper, liquid structure is computed by solving integral equations based on the Ornstein-Zernike (OZ) equation \cite{Hansen_McDonald1986}.
This approach requires an approximative closure for an explicit solution. A rather simple but successful closure is the
hypernetted chain (HNC) scheme \cite{Morita1958}, which has been proven to be a realistic approximation for mixtures of charged particles.
The numerical solution methods presented here can be straightforwardly generalized to more sophisticated,
thermodynamically partially self-consistent OZ closure relations \cite{Caccamo1996, Nagele1996, Caccamo1999}
like, e.g., the one proposed by Zerah and Hansen \cite{Zerah1986}.
For the sake of simplicity, and since it was shown that enforcing thermodynamic self-consistency leads
to a weak accuracy improvement only \cite{Belloni1988}, in the present work we restrict ourselves to the HNC approximation. 
A variety of liquid integral equation \cite{Khan1987, Gonzalez-Mozuelos1998, Carbajal-Tinoco2002, Belloni1988, Belloni2002, Anta2005, Camargo2008}
or density functional studies \cite{Lowen1992, Fushiki1992, Lowen1993} of the PM have been reported.
The HNC equations have been solved for the asymmetric PM by L\'{e}ger and Levesque \cite{Leger2005}, in case of
non-zero macroion number densities for size asymmetries between micro- and macroions as high as 1:80 and charge asymmetries
ranging up to 1:450.

The HNC scheme can be formulated in any spatial dimension $d>0$. While $d=3$ is the standard three-dimensional situation,
it is important to note that also two-dimensional and one-dimensional fluids occur in experiments with strong
confinement between glass plates or at interfaces, or in one-dimensional channels \cite{Lowen2001}.
Dimensions higher than $d=3$ have no immediate realization. However, they play an important role in constructing
and testing theories and are also helpful to find suitable mean-field-like approximations in lower dimensions
\cite{Finken2002, vanMeel2009a, vanMeel2009b, Charbonneau2011}.
Hence, there is a need to study charged systems also in $d>3$. We formulate the solution method in the present paper
for arbitrary $d>0$, with explicit data presented for spatial dimensions $d=1,2,3,4,5,$ and $6$.

Here we pick up a strategy of solving the PM-HNC equations that, for $d=3$, has first been employed
by Rossky and Friedman \cite{Rossky1980}. The key idea lies in using computational grids with logarithmic spacing
in coordinate- and wavenumber space. For even number of dimensions $d$, in particular for $d=2$,
the use of logarithmic grids in a spectral OZ equation solver emerges as a necessary consequence of mapping the occurring
Fourier-Bessel transforms to numerically efficient fast Fourier transform (FFT) methods \cite{Gardner1959, Siegman1977, Talman1978, Hamilton2000, Hamilton_website}.
Hence, logarithmic grids appear quite naturally in $d=2$ liquid integral equation studies like, \textit{e.g.},
\refsname{}\cite{Caillol1981, Weis1998}.
For odd number of dimensions, where FFT methods can be applied directly on uniformly spaced grids,
using logarithmic grids is a less obvious approach which has nevertheless been employed in some $d=3$
studies \cite{Kalyuzhnyi2001, Lin2002, Kloss2008}. In none of these studies, however,
extensive use has been made of an important virtue of the logarithmic grids, which is highlighted in our present paper:  
The simultaneous dense distribution of grid points at very different length scales renders logarithmic grids
ideal for the discretization of pair-correlation functions in a highly asymmetric PM, at moderate numerical expense.
Here we are presenting results for ion size- and charge-asymmetries as high as 1:1000
(both asymmetries simultaneously reached, for non-dilute suspensions), which represents well the colloidal regime. To our knowledge, no liquid
integral equation or computer simulation studies have been published so far, where asymmetries of this magnitude have been reached.

We investigate the accuracy of the HNC pair-correlation functions by comparing to results of numerically expensive
Molecular Dynamics (MD) simulations, for ion charge- and diameter asymmetries up to 1:500 and 1:250, respectively.
The HNC results are found to be in overall good agreement with the MD results, except for a somewhat underestimated
principal peak in the macroion radial distribution function at high macroion charge numbers.

Our results can be used to extract colloidal effective interaction potentials, including non-saturated effective charge numbers, at high numerical efficiency.
As opposed to other approximate theories for colloidal effective charge numbers like the cell model \cite{Alexander1984, Allahyarov2005} or the
renormalized jellium model \cite{Trizac2004, Colla2009}, pair correlations among all ionic species are treated on equal footing in our method,
with the HNC entering as the only approximation. In a future study, the method described here could be augmented to
include colloidal surface chemistry, described by a mass action balance that takes into account the local variations in the pH value
near the colloidal surfaces. This would allow for parameter free \textit{ab initio} calculation of pair correlations in colloidal
suspensions with reactive electrolyte like, \textit{e.g.}, suspensions of silica spheres in $NaOH$ with a reentrant fluid-crystal-fluid
phase diagram, reported in \refsname\cite{Herlach2010, Wette2010}.

% This paper is organized as follows:
% In \sectionname \ref{sec:sub:PM}, the PM of an ionic fluid mixture in arbitrary spatial dimension is laid out.
% Description of our numerical and computer simulation methods commences with the formulation of the HNC equations in 
% \sectionname \ref{sec:sub:HNC}. In the following \sectionname\ref{sec:sub:Algo} we give a detailed explanation of our numeric method to solve the HNC equations.
% The Fourier-transform on logarithmic grids, which is at the heart of our solution method, is discussed in \sectionname \ref{sec:sub:Log}.
% Both \sectionname\ref{sec:sub:Algo} and \ref{sec:sub:Log} are written in a technical, self-contained style, to ensure reproducibility of the numerical methods.
% In \sectionname\ref{sec:sub:MD}, the technical aspects of our MD simulations are briefly mentioned. 
% The results section begins with a discussion of thermodynamic properties and pair-correlation functions 
% of the PM in up to six spatial dimensions in \sectionname{}\ref{sec:sub:Dim1to6}. In the following \sectionname{}\ref{sec:sub:Coll}, we apply our method to
% three-dimensional suspensions of micron-sized charged colloidal spheres in 1-1 electrolytes.
% Effective interaction potentials between colloidal particles are extracted from the full primitive model solution in
% \sectionname \ref{sec:sub:Eff_int}, which includes a discussion of effective colloidal charge numbers.
%%%%%%%%%%%%%%%%%%%%%%%%%%%%%%%%%%%%%%%%%%%%%%%%%%%%%%%%%%%%%%%%%%%%%%%%%%%%%%%%%%%%%%%%%%%%%%%%%%%%%%%%%%%%%%%%

\section{Methods}\label{sec:Methods}

%%%%%%%%%%%%%%%%%%%%%%%%%%%%%%%%%%%%%%%%%%%%%%%%%%%%%%%%%%%%%%%%%%%%%%%%%%%%%%%%%%%%%%%%%%%%%%%%%%%%%%%%%%%%%%%%
\subsection{The $\boldsymbol{d}$-dimensional primitive model}\label{sec:sub:PM}

Let $n_i$, with $1 \leq i \leq m$, denote the number density of ions of species $i$ in an $m$-component fluid
mixture of hyperspherical particles in an arbitrary positive number $d$ of spatial dimensions.
Then, $n = \sum_i n_i$ is the total particle number density and $\chi_i = n_i / n$ is the mole fraction
of species $i$. Let $Z_i e$ denote the electric charge of a particle of species $i$, where $e$ is the elementary charge.
In order to prevent singular attractions between oppositely charged particles, ions of species $i$ possess a hard-core diameter $\sigma_i$.
Hence, the particles of species $i$ occupy a fraction
\begin{equation}\label{equ:hypervolfrac}
\phi_i = V(d) (\sigma_i / 2)^{d} n_i 
\end{equation}
of the system hypervolume,
where $V(d) = \pi^{d/2}/\Gamma(d/2+1)$ is the $d$-dimensional unit hypersphere volume, $\Gamma(x)$ denoting the Gamma function.
All ions are assumed neutrally buoyant in an infinite structureless solvent, which is fully described by the solvent
dielectric constant $\epsilon$ in the PM applied here.
We express all real-space functions in units of the dimensionless particle-center to particle-center distance $x = r n^{1/d}$.

The functions  $u_{ij}(x) = V_{ij}(x) / (k_B T)$ are the pair-potentials of direct interaction, $V_{ij}(x)$, 
between ions of species $i$ and $j$, divided by the Boltzmann constant $k_B$ and the absolute temperature $T$.
The dimensionless pair-potentials can be decomposed as 
\begin{equation} \label{equ:pair_pot}
u_{ij}(x) = u_{ij}^{(s)}(x) + u_{ij}^{(l)}(x)  
\end{equation}
into short-ranged (hard-core) parts
\begin{equation} \label{equ:u_hardcore}
u_{ij}^{(s)}(x) = \left\lbrace
   \begin{array}{ll}
   \infty \,\rule[-1.5em]{0em}{1em} & ~~\text{for}~x < \sigma_{ij} n^{1/d},\\
   0 & ~~\text{otherwise,}%\rule[-1.4em]{0em}{.9em},
   \end{array}
 \right. \, \\
\end{equation}
with pairwise additive hard-core diameters $\sigma_{ij} = 1/2 (\sigma_i + \sigma_j)$,
and long-ranged (Coulomb) parts
\begin{equation} \label{equ:u_Coulomb}
u_{ij}^{(l)}(x) = \left\lbrace
   \begin{array}{ll}
   -\Gamma_{ij} \ln(x) \,& ~~\text{for}~d = 2,\\~\\
   \frac{\displaystyle{\Gamma_{ij}}\rule{0em}{1em}}{\displaystyle{(d-2) x^{d-2}}\rule{0em}{.9em}}\,& ~~\text{for}~d \neq 2,\\
   \end{array}
 \right. \, \\
\end{equation}
with coupling constants $\Gamma_{ij} \propto Z_i Z_j$. 
In $d=3$ dimensions, connection to experimentally accessible systems can be made by choosing the coupling constant $\Gamma_{ij}$
in \expressionname\eqref{equ:u_Coulomb} as \mbox{$\Gamma_{ij} = L_B n^{1/3} Z_{i} Z_{j}$}, involving the solvent-specific Bjerrum
length \mbox{$L_B = e^2/(\epsilon k_B T)$} in Gaussian units.
In the present paper, we investigate only such systems that obey the Berthelot mixing rule $\Gamma_{ij}^2 = \Gamma_{ii}\Gamma_{jj}$ \cite{Hopkins2006}.

One-component systems, $m=1$, with pair-potential according to \expressionsname\eqref{equ:pair_pot}-\eqref{equ:u_Coulomb},
are commonly referred to as ($d$-dimensional) one component plasmas (OCPs). Global electroneutrality of an OCP
implies the presence of a homogeneous background charge density that does not couple to the distribution of the correlated ions
(like, \textit{e.g.}, an electron plasma at sufficiently high temperature). For systems with $m>1$ components, global electroneutrality without a
neutralizing charge background is enforced in all cases studied here, by requiring that $\sum_i n_i Z_i = 0$.

\subsection{Hypernetted chain scheme}\label{sec:sub:HNC}

We compute the ion pair-correlations in the PM described in the previous section, by numerically solving the
OZ equations \cite{Hansen_McDonald1986} in combination with the approximate HNC closure \cite{Morita1958, Caccamo1996, Nagele1996}. 
In an isotropic, homogeneous fluid mixture, the coupled OZ equations may be written as
\begin{equation}\label{equ:OZ_unit_r}
\gamma_{ij}(x) = \chi_k \int d^d \mathbf{x'} c_{ik}(x')\left[c_{kj}(x-x') + \gamma_{kj}(x-x')\right]. 
\end{equation}
In \expressionsname\eqref{equ:OZ_unit_r} and the rest of this paper we adhere to the Einstein summation convention.
Equations \eqref{equ:OZ_unit_r} can be regarded as the definitions of the partial direct correlation functions $c_{ij}(x)$ in terms of
the continuous partial indirect correlation functions $\gamma_{ij}(x) = h_{ij}(x) - c_{ij}(x) = g_{ij}(x) - 1 -c_{ij}(x)$. The latter
identity comprises the total correlation functions $h_{ij}(x)$ and the
partial radial distribution functions (rdf's), $g_{ij}(x)$, which are the conditional probabilities of
finding a particle of species $j$ at a dimensionless center-to-center distance $x$ from a particle of species $i$ \cite{Hansen_McDonald1986}.

An isotropic function $f$ can be Fourier-transformed in $d$ dimensions as
\begin{eqnarray}
\tilde{f}(y) &=& \frac{{(2\pi)}^{d/2}}{y^{d/2-1}} \int\limits_0^\infty dx~ x^{d/2} f(x) J_{d/2-1}(xy), \label{equ:Fourier_x_to_y}\\
{f}(x) &=& \frac{x^{1-d/2}}{{(2\pi)}^{d/2}} \int\limits_0^\infty dy~ y^{d/2} \tilde{f}(y) J_{d/2-1}(xy), \label{equ:Fourier_y_to_x}
\end{eqnarray}
with $J_{n}(x)$ denoting the Bessel function of the first kind and order $n$. Employing the convolution theorem,
the OZ equations are Fourier-transformed into the space of dimensionless wavenumbers $y$, where they read
\begin{equation}\label{equ:OZ_yspace}
\tilde{\gamma}_{ij}(y) = \chi_k \tilde{c}_{ik}(y) \tilde{c}_{kj}(y) + \chi_k \tilde{c}_{ik}(y) \tilde{\gamma}_{kj}(y).   
\end{equation}

The OZ equations need to be supplemented by an appropriate closure relation.
The HNC closure, which is known to be a good approximation for the PM \cite{Belloni1988}, reads
\begin{equation}\label{equ:HNC_unit_r}
c_{ij}(x) = \exp\{ \gamma_{ij}(x) - u_{ij}(x) \} - \gamma_{ij}(x) - 1.
%\gamma_{ij}(x) = \ln\left[g_{ij}(x)\right] + u_{ij}(x).
\end{equation}

Numerical solution of the set of \expressionsname \eqref{equ:OZ_unit_r} and \eqref{equ:HNC_unit_r} in combination with the long-ranged
potentials in \expressionsname \eqref{equ:pair_pot}-\eqref{equ:u_Coulomb} requires splitting the analytically known long-ranged asymptotic parts
$\mp u_{ij}^{(l)}(x)$ off the direct and indirect correlation functions as \cite{Ng1974, Leger2005, Hansen_McDonald1986} 
\begin{equation}\label{equ:c_short_long}
c_{ij}(x) = c_{ij}^{(s)}(x) + u_{ij}^{(l)}(x)
\end{equation}
and
\begin{equation}\label{equ:gamma_short_long}
\gamma_{ij}(r) = \gamma_{ij}^{(s)}(x) - u_{ij}^{(l)}(x).
\end{equation}
The so-defined functions $c_{ij}^{(s)}(x)$ and $\gamma_{ij}^{(s)}(x)$ are considerably shorter in range than
$c_{ij}(x)$ and $\gamma_{ij}(x)$.
In terms of the short-ranged correlation functions and long-ranged potential parts, the OZ equations in wavenumber space can be written as
the set of coupled algebraic equations
\begin{equation}\label{equ:OZ_short}
\left[\delta_{ik} - \chi_k \tilde{c}_{ik}^{(s)} + \chi_k \tilde{u}_{ik}^{(l)} \right] \tilde{\gamma}_{kj}^{(s)} = 
-\tilde{u}_{ij}^{(l)} - \chi_k \tilde{u}_{ik}^{(l)} \tilde{c}_{kj}^{(s)} + \chi_k \tilde{c}_{ik}^{(s)} \tilde{c}_{kj}^{(s)} 
\end{equation}
with Kronecker-delta $\delta_{ij}$.
The HNC closure in terms of the short-ranged correlation functions and the hard-core diameters is 
\begin{equation}\label{equ:HNC_short}
c_{ij}^{(s)}(x) = \Theta(x-\sigma_{ij}n^{1/d})\exp\{\gamma_{ij}^{(s)}(x)\} - \gamma_{ij}^{(s)}(x) - 1,
\end{equation}
with unit step function $\Theta(x)$.

The Fourier transform of the Coulombic part $u_{ij}^{(l)}(x)$ of the potential is \cite{Oberhettinger}
\begin{equation}\label{equ:u_trans_long}
\tilde{u}_{ij}^{(l)}(y) = 
   \frac{\displaystyle{\Gamma_{ij}A(d)}}{\displaystyle{y^2}\rule{0em}{1em}},
\end{equation}
where $A(d) = 2 \pi^{d/2} / \Gamma(d/2)$ denotes the surface of the $d$-dimensional unit hypersphere.

We solve the closed set of \expressionsname\eqref{equ:OZ_short}-\eqref{equ:u_trans_long} by the numeric methods described in the
following two sections. Our results are presented in form of
the partial rdf's $g_{ij}(x)$, and partial static structure factors, $S_{ij}(y) = \delta_{ij} + \sqrt{\chi_i \chi_j} \tilde{h}_{ij}(y)$
\cite{Hansen_McDonald1986, Nagele1996}.

\subsection{Numerical algorithm}\label{sec:sub:Algo}

In order to solve the set of \expressionsname\eqref{equ:OZ_short}-\eqref{equ:u_trans_long}, we employ a generalized version of 
the numerically robust, quickly convergent solution method introduced by Ng in the appendix of \refname{}\cite{Ng1974}. The method
shares great similarities with the direct inversion of iterative subspace (DIIS) method developed by Pulay\cite{Pulay1980, Pulay1982},
which is commonly used in the solution of quantum mechanical (Hartree-Fock) 
density functional problems (see \refname{}\cite{Rohwedder2011} for a detailed analysis of the DIIS method). The DIIS method has been applied also to
density functional theory of hard spheres \cite{Haertel2012}, and reference interaction site model HNC equations of liquid water \cite{Kovalenko1999}. 

In conformity with Ng's notation, we formulate a fixed point problem
\begin{equation}\label{equ:fixed_point_problem}
\boldsymbol{A} \cdot \boldsymbol{c}^{(s)}(x) \stackrel{!}{=} \boldsymbol{c}^{(s)}(x),   
\end{equation}
to be fulfilled by the exact solutions $\boldsymbol{c}^{(s)}(x)$ of \expressionsname\eqref{equ:OZ_short}-\eqref{equ:u_trans_long}, 
for arbitrary values of the coordinate $x$.
Equation \eqref{equ:fixed_point_problem} contains the $m \times m$ function array $\boldsymbol{c}^{(s)}(x)$ with elements $c_{ij}^{(s)}(x)$, and the
nonlinear fourth-rank operator $\boldsymbol{A}$, that depends on all pair potentials $u_{ij}(x)$.

We solve \expressionname\eqref{equ:fixed_point_problem} numerically by executing two nested instruction loops
with iteration indices \mbox{$n_1 \in \{ \mathbb{N} \cap [0,n_1^{\text{max}}] \}$} and \mbox{$n_2 \in \mathbb{N}$}, such that
$\lim_{n_2\to\infty} \boldsymbol{c}^{(s)}_{\{n_1^{\text{max}},n_2\}}(x) = \boldsymbol{c}^{(s)}(x)$,
provided that the iteration with respect to index $n_2$ converges.
To avoid confusion with other kinds of indices or with particle number densities in the following, $n_1$ and $n_2$ are exclusively used
to label entire operators, matrices or vectors (identified through bold font), and are never used in indexing scalars.
Both $n_1$ and $n_2$ are always enclosed in curly brackets when used as an index, whereas other lower indices, like the species indices
$i, j, \ldots$ stand always without brackets.
For instance, $\boldsymbol{c}^{(s)}_{\{n_1,n_2\}}(x)$ is the $m \times m$ array of intermediate solutions for the short-ranged parts of partial
direct correlation functions at iteration stage characterized by $n_1$ and $n_2$, and the element of that array with particle-species
indices $i$ and $j$ is the function ${(\boldsymbol{c}^{(s)}_{\{n_1,n_2\}})}_{ij}(x)$.

In the outer loop with index $n_1$, the elements of an $m \times m$ matrix $\boldsymbol{\Gamma}_{\{n_1\}}$ of coupling parameters
and an $m$-dimensional vector $\boldsymbol{\phi}_{\{n_1\}}$ of hypervolume fractions are both ramped up from
initial values ${(\boldsymbol{\Gamma}_{\{0\}})}_{ij}$ and ${(\boldsymbol{\phi}_{\{0\}})}_{i}$ of small magnitude,
to their final values ${(\boldsymbol{\Gamma}_{\{n_1^{\text{max}}\}})}_{ij} = \Gamma_{ij}$ and ${(\boldsymbol{\phi}_{\{n_1^{\text{max}}\}})}_{i} = \phi_{i}$,
characterizing the pair-potentials to be solved for through \expressionsname\eqref{equ:hypervolfrac}-\eqref{equ:u_Coulomb}.
We employ the rules
\begin{equation}\label{equ:coupling_rampup}
\boldsymbol{\Gamma}_{\{n_1\}} = \varepsilon(n_1) \boldsymbol{\Gamma}_{\{n_1^{\text{max}}\}}
\end{equation}
and
\begin{equation}\label{equ:hypervolfrac_rampup}
\boldsymbol{\phi}_{\{n_1\}} = {\left[\varepsilon(n_1)\right]}^{1/10} \boldsymbol{\phi}_{\{n_1^{\text{max}}\}}
\end{equation}
for potential ramp-up, where a near-optimal convergence rate, combined with good numerical stability of the outer loop iteration
is achieved by a convergence-adaptive scaling parameter $0 < \varepsilon \leq 1$, which increases monotonically as a function of $n_1$.
After each outer iteration, the growth rate of $\varepsilon(n_1)$ is increased if the previous inner loop took less than a certain
threshold of iterations to converge, and decreased in the opposite case.

Our experience shows that numerical stability of the algorithm benefits considerably from 
the superlinear form of \expressionname\eqref{equ:hypervolfrac_rampup}, characterized by the (empirically chosen) exponent $1/10$.
This can be rationalized by considering the contact value, $\lim_{x\searrow\sigma_{ij}n^{1/d}}u_{ij}(x)$, of the pair-potential in
\expressionsname\eqref{equ:hypervolfrac}-\eqref{equ:u_Coulomb}: The contact value decreases for increasing values of $\phi_i$ and $\phi_j$,
and increases for increasing $\Gamma_{ij}$. The potential at particle contact has great influence on the strength of the undulations in the
pair-structure functions, which, in turn, influence numerical stability.
Therefore, it is favorable to choose potential ramp-up rules like \expressionsname\eqref{equ:coupling_rampup}
and \eqref{equ:hypervolfrac_rampup}, where the $\phi_i$ are increased quicker than the $\Gamma_{ij}$.

That a potential ramp-up is necessary at all,
is owed to critical dependence of the inner loop convergence on the quality of the inner loop seed
$\boldsymbol{c}_{\{n_1,0\}}^{(s)}$. A good analytical estimate of $\boldsymbol{c}_{\{n_1,0\}}^{(s)}$ exists only for low magnitudes of all
${(\boldsymbol{\Gamma}_{\{n_1\}})}_{ij}$ and ${(\boldsymbol{\phi}_{\{n_1\}})}_{i}$. The inner iteration seeds
used in our algorithm are given in \expressionsname\eqref{equ:Ng_iter_seed_0}-\eqref{equ:Ng_iter_seed_2}, and 
rationalized in the surrounding text.

We proceed now with the discussion of the inner iteration loop, where $n_1$ is kept fixed.
Operator $\boldsymbol{A}_{\{n_1\}}$ is defined by
\begin{equation}\label{equ:A_operator_definition}
\boldsymbol{A}_{\{n_1\}} \cdot \boldsymbol{c}^{(s)}_{\{n_1,n_2\}}(x) = \boldsymbol{\eta}^{(s)}_{\{n_1,n_2\}}(x),   
\end{equation}
as the operator that transforms the input functions arrays, $\boldsymbol{c}^{(s)}_{\{n_1,n_2\}}(x)$, for fixed indices $n_1$ and $n_2$,
into the corresponding output functions arrays, $\boldsymbol{\eta}_{\{n_1, n_2\}}(x)$, the latter being defined further down the text of this subsection.
Hence, $\boldsymbol{A}_{\{n_1^{\text{max}}\}} = \boldsymbol{A}$, and equation \eqref{equ:fixed_point_problem} is equivalent to
\begin{equation}\label{equ:Ng_difference_arrays_criterion}
\lim\limits_{n_2\to\infty} \boldsymbol{d}_{\{n_1^{\text{max}},n_2\}}(x) \stackrel{!}{=} 0,
\end{equation}
with function arrays $\boldsymbol{d}_{\{n_1,n_2\}}(x)$ defined by
\begin{eqnarray}
\boldsymbol{d}_{\{n_1,n_2\}}(x) &&= \boldsymbol{\eta}_{\{n_1,n_2\}}(x) - \boldsymbol{c}^{(s)}_{\{n_1,n_2\}}(x)\nonumber\\
&&= \left(\boldsymbol{A}_{\{n_1\}} - \mathds{1}\right)\cdot\boldsymbol{c}^{(s)}_{\{n_1,n_2\}}(x).\label{equ:Ng_difference_arrays_definition}
\end{eqnarray}

In our implementation, iteration in the inner loop 
is stopped at a finite, $n_1$-dependent value, $n_2^{\text{max}}(n_1) > 1$, of the index $n_2$, once the convergence criterion 
\begin{equation}\label{equ:conv_crit}
\frac{\displaystyle{
\| \boldsymbol{d}_{\{n_1,n_2^{\text{max}}(n_1)\}}(x) \|}}
{\displaystyle{
\| \boldsymbol{\eta}_{\{n_1,n_2^{\text{max}}(n_1)\}}(x) \|}}
< \text{TOL}(n_1), 
\end{equation}
with a small tolerance TOL$(n_1)$, as specified below, has been fulfilled.
In \expressionname\eqref{equ:conv_crit}, the norm, $\|\textbf{f}(x)\|$, of an $m \times m$ function array $\boldsymbol{f}(x)$,
is defined as
\begin{equation}\label{equ:norm_definition}
\|\textbf{f}(x)\| = {\left(\boldsymbol{f}(x), \boldsymbol{f}(x)\rule[-.1em]{0em}{1.2em}\right)}^{1/2}, 
\end{equation}
and the bracket $\left(\boldsymbol{f}, \boldsymbol{g}\right)$ denotes the inner product
\begin{equation}\label{equ:inner_prod}
\left(\boldsymbol{f}, \boldsymbol{g} \right) = 
\int\limits_{x_1}^{x_2} \boldsymbol{f}(x) : \boldsymbol{g}(x) dx, 
\end{equation}
of two $m \times m$ function arrays $\boldsymbol{f}(x)$ and $\boldsymbol{g}(x)$ with elements $f_{ij}(x)$ and $g_{ij}(x)$,
double dots indicating the contraction with respect to both particle species indices $i$ and $j$.
The interval $[x_1, x_2]$ should be chosen to contain the major structural features of the partial direct correlation functions.
To obtain the results presented in this paper, we have used $x_1 = \min\{0.5 \sigma_i n^{1/d}, i=1\ldots m\}$,
$x_2 = 30$, and values of TOL that decrease as a function of $n_1$, with \mbox{TOL$(0) < 10^{-4}$} and \mbox{TOL$(n_1^{\text{max}}) < 10^{-12}$}.

In our algorithm, $\boldsymbol{\eta}_{\{n_1, n_2\}}(x)$ is obtained from $\boldsymbol{c}^{(s)}_{\{n_1, n_2\}}(x)$ at fixed values of $n_1$ and $n_2$
by applying the following four steps:

\begin{center}
\emph{Step 1:}
\end{center}
The function arrays $\boldsymbol{c}^{(s)}_{\{n_1, n_2\}}(x)$ are Fourier-transformed into $\tilde{\boldsymbol{c}}^{(s)}_{\{n_1, n_2\}}(y)$
by the fast transform method described in the following section, requiring logarithmically spaced grids in $x$- and $y$-space.

\begin{center}
\emph{Step 2:}
\end{center}
The coupled OZ \expressionsname \eqref{equ:OZ_short} are solved to obtain $\tilde{\boldsymbol{\gamma}}^{(s)}_{\{n_1, n_2\}}(y)$ from 
$\tilde{\boldsymbol{c}}^{(s)}_{\{n_1, n_2\}}(y)$.

\begin{center}
\emph{Step 3:}
\end{center}
The fast inverse transform method on logarithmic grids is applied to compute the inverse
Fourier transform, $\boldsymbol{\gamma}^{(s)}_{\{n_1, n_2\}}(x)$,
of the function array $\tilde{\boldsymbol{\gamma}}^{(s)}_{\{n_1, n_2\}}(y)$.

\begin{center}
\emph{Step 4:}
\end{center}
The elements of the function arrays $\boldsymbol{\eta}_{\{n_1, n_2\}}(x)$ are calculated as the left-hand-sides
of \expressionsname\eqref{equ:HNC_short} (the HNC closure), where the elements of the function arrays ${\boldsymbol{\gamma}}^{(s)}_{\{n_1, n_2\}}(x)$
from step 3 are entered to the right-hand-sides of \expressionsname\eqref{equ:HNC_short}.\\~\\

A straightforward way of selecting the input functions, $\boldsymbol{c}^{(s)}_{\{n_1, n_2+1\}}(x)$, for the next step of the inner loop,
is the Picard-iteration scheme
\begin{equation}\label{equ:Picard}
\boldsymbol{c}^{(s)}_{\{n_1,n_2+1\}}(x) = \boldsymbol{\eta}_{\{n_1, n_2\}}(x). 
\end{equation}
This simple scheme, however, converges only for weak pair-potentials (small values of $|\Gamma_{ij}|$ and $\phi_i$, in our case).

Numerical stability of the Picard-iteration scheme can be somewhat improved, at the cost of increasing computational effort,
if a fixed mixing parameter $0 < \alpha < 1$ is introduced in \expressionname\eqref{equ:Picard},
which gives the alternative iteration rule 
\begin{equation}\label{equ:mixing_scheme}
\boldsymbol{c}^{(s)}_{\{n_1,n_2+1\}}(x) = \alpha \boldsymbol{\eta}_{\{n_1,n_2\}}(x) + (1-\alpha) \boldsymbol{c}^{(s)}_{\{n_1,n_2\}}(x).   
\end{equation}
Fixed point iterations on basis of \expressionname\eqref{equ:Picard} or \eqref{equ:mixing_scheme}
have been applied in a number of integral equation studies\cite{Patey1977, Rossky1980, Caillol1981, Weis1998, Leger2005, Hoffmann2007, Camargo2008},
where different strategies have been applied in computing the Fourier transforms, and various closure relations for the OZ-equations have been used,
including the HNC closure. The mixing parameter $\alpha$ in \expressionname\eqref{equ:mixing_scheme} has been empirically determined in most cases.

Despite being numerically more robust than the Picard iteration in \expressionname\eqref{equ:Picard},
the iteration scheme according to \expressionname\eqref{equ:mixing_scheme} still fails to converge for large values
of $|\Gamma_{ij}|$ or $\phi_i$, especially if the number of components, $m$,
is larger than one. We therefore use a generalized version of the fixed point iteration scheme proposed by Ng \cite{Ng1974},
which has proven to be numerically much more stable and efficient. Ng's iteration scheme, generalized to
$m$-component mixtures and arbitrary number, $M\geq0$, of mixing coefficients, reads
\begin{eqnarray}
\boldsymbol{c}^{(s)}_{\{n_1,n_2+1\}}(x) = \left(1 - \sum\limits_{l=1}^{M} {\left(\boldsymbol{\alpha}_{\{n_1,n_2\}}\right)}_{l}\right)\boldsymbol{\eta}_{\{n_1,n_2\}}(x)\nonumber\\
+\sum\limits_{l=1}^{M} {\left(\boldsymbol{\alpha}_{\{n_1,n_2\}}\right)}_{l} \boldsymbol{\eta}_{\{n_1,n_2-l\}}(x).\nonumber\\ \label{equ:Ng_scheme}
\end{eqnarray}
For $M=0$, \expressionname\eqref{equ:Ng_scheme} reduces to the Picard iteration in \expressionname\eqref{equ:Picard}.
At every step of the iteration, the $M$-dimensional mixing coefficient vector $\boldsymbol{\alpha}_{\{n_1,n_2\}}$ is
determined as the solution of the set of linear equations
\begin{equation}\label{equ:Ng_mixcoeffs_LGS}
\boldsymbol{\Delta}_{\{n_1,n_2\}} \cdot \boldsymbol{\alpha}_{\{n_1,n_2\}} = \boldsymbol{\delta}_{\{n_1,n_2\}}, 
\end{equation}
where the elements, 
\begin{equation}
{\left(\boldsymbol{\Delta}_{\{n_1,n_2\}}\right)}_{lm} =
\left( \boldsymbol{v}_{\{n_1,n_2,l\}}, \boldsymbol{v}_{\{n_1,n_2,m\}} \right)\label{equ:Ng_mixcoeffs_coeff_matrix}\\  
\end{equation}
and
\begin{equation}
{\left(\boldsymbol{\delta}_{\{n_1,n_2\}}\right)}_l = \left( \boldsymbol{d}_{\{n_1,n_2\}}, \boldsymbol{v}_{\{n_1,n_2,l\}} \right),\label{equ:Ng_mixcoeffs_LGS_rhs}
\end{equation}
of the $M \times M$ matrix $\boldsymbol{\Delta}_{\{n_1,n_2\}}$ and the vector $\boldsymbol{\delta}_{\{n_1,n_2\}}$
are inner products involving the function arrays $\boldsymbol{d}_{\{n_1,n_2\}}$ and 
\begin{equation}\label{equ:Ng_difference_arrays_tripleindex}
\boldsymbol{v}_{\{n_1,n_2,l\}} = \boldsymbol{d}_{\{n_1,n_2\}} - \boldsymbol{d}_{\{n_1,n_2-l\}}.
\end{equation}

Solving for $\boldsymbol{\alpha}_{\{n_1, n_2\}}$ in \expressionsname\eqref{equ:Ng_mixcoeffs_LGS}-\eqref{equ:Ng_difference_arrays_tripleindex}
is equivalent to solving the minimization problem
\begin{equation}\label{equ:Ng_minimized_norm}
\left\| \boldsymbol{d}_{\{n_1,n_2\}} - \sum\limits_{l=1}^M {\left(\boldsymbol{\alpha}_{\{n_1,n_2\}}\right)}_l
\boldsymbol{v}_{\{n_1,n_2,l\}} \right\| \stackrel{!}{=} \text{min},
\end{equation}
with respect to $\boldsymbol{\alpha}_{\{n_1,n_2\}}$, and the minimization in expression \eqref{equ:Ng_minimized_norm} can be
motivated by approximating $\boldsymbol{A}_{\{n_1\}}$ as a locally linear operator \cite{Ng1974, Mendez-Alcaraz-Thesis}.
%For a more detailed derivation of \expressionsname\eqref{equ:Ng_scheme}-\eqref{equ:Ng_minimized_norm},
%formulated for the special case of $m=1$ and $M=2$, we refer to Ng's original article in \refname{}\cite{Ng1974}.

It may be counterintuitive, but is worthwhile to note that one should \textit{not} select $M=n_2$, \textit{i.e.}, the maximum possible order at each
inner loop iteration. Instead, numerical stability is increased if one chooses $M$ to rise slower than possible,
in our case as
\begin{equation} \label{equ:Ng_order_increase}
M(n_2) = \left\lbrace
   \begin{array}{ll}
   \text{min}\{n_2 ,M^{\text{max}}\}\,\rule[-1.5em]{0em}{1em} & ~~\text{for}~n_2 \leq 5,\\
   \text{min}\{2 + \lfloor n_2/2 \rfloor, M^{\text{max}}\} & ~~\text{for}~n_2>5,
   \end{array}
 \right. \, \\
\end{equation}
with $\lfloor a \rfloor$ denoting the largest integer number smaller than or equal to $a$, 
and $M^{\text{max}} = 20$, which results in swift convergence.
Presumably, the reason for the increased numerical stability of the rule in \expressionname\eqref{equ:Ng_order_increase} as compared to the rule $M=n_2$, is that the
low-quality intermediate solutions for small values of $n_2$ are always retained in the mixing rule in \expressionname\eqref{equ:Ng_scheme} if $M=n_2$ is chosen,
whereas they are dismissed, and thereby prevented from spoiling convergence at sufficiently high $n_2$, if $M$ rises more slowly than $n_2$.

The iteration scheme defined by \expressionsname\eqref{equ:Ng_scheme}-\eqref{equ:Ng_difference_arrays_tripleindex}
fails to converge, if the seed of the inner loop iteration,
$\boldsymbol{c}_{\{n_1, 0\}}^{(s)}(x)$, is too different from the fixed point of operator $\mathbf{A}_{\{n_1\}}$. 
A good analytical estimate of an iteration seed exists only for sufficiently small
coupling parameters and hypervolume fractions. In this regime, one may approximate the $c_{ij}(x)$ by their infinite dilution (number density $n\to0$) limit
$c_{ij}(x) \to f_{ij}(x) = \exp\{-u_{ij}(x)\} - 1$, where the $f_{ij}(x)$ are the Mayer functions \cite{Hansen_McDonald1986, Hoffmann2007}.
This results in a seed
\begin{equation}\label{equ:Ng_iter_seed_0}
\boldsymbol{c}_{\{0,0\}}^{(s)}(x) = 
   \exp[-\boldsymbol{u}_{\{0\}}(x)] - 1 + \boldsymbol{u}_{\{0\}}(x)
\end{equation}
for $n_1 = n_2 = 0$, where $\boldsymbol{u}_{\{n_1\}}(x)$ denotes
an array of pair-potentials ${(\boldsymbol{u}_{\{n_1\}})}_{ij}(x)$
between particles of species $i$ and $j$, obtained from inserting the 
potential parameters ${(\boldsymbol{\Gamma}_{\{n_1\}})}_{ij}$ and ${(\boldsymbol{\phi}_{\{n_1\}})}_i$
into \expressionsname\eqref{equ:hypervolfrac}-\eqref{equ:u_Coulomb}.

To access HNC solutions at higher $|\Gamma_{ij}|$ and $\phi_i$, we construct the seeds for $n_1 > 0$
from the converged solutions of inner iterations corresponding to smaller values of $n_1$. For $n_1 = 1$, we choose
\begin{equation}\label{equ:Ng_iter_seed_1}
\boldsymbol{c}_{\{1,0\}}^{(s)}(x) = \boldsymbol{\eta}_{\{0,n_2^{\text{max}}(0)\}}(x),
\end{equation}
which is a Picard-iteration step in the outer loop.
For $n_1 > 1$, we use the seed
\begin{eqnarray}\label{equ:Ng_iter_seed_2}
\boldsymbol{c}_{\{n_1,0\}}^{(s)}(x) =&& s^{(1)} \boldsymbol{\eta}_{\{n_1-1,n_2^{\text{max}}(n_1-1)\}}(x),\nonumber\\
+&& s^{(2)} \boldsymbol{\eta}_{\{n_1-2,n_2^{\text{max}}(n_1-2)\}}(x),				  
\end{eqnarray}
with coefficients
\begin{equation}
s^{(1)} =\frac{\varepsilon(n_1) - \varepsilon(n_1-2)}{\varepsilon(n_1-1) - \varepsilon(n_1-2)} 
\end{equation}
and
\begin{equation}\label{equ:extrap_coeff2}
s^{(2)} =\frac{\varepsilon(n_1) - \varepsilon(n_1-1)}{\varepsilon(n_1-1) - \varepsilon(n_1-2)},
\end{equation}
that extrapolate linearly on basis of the previous two converged inner iteration solutions. While the linear extrapolation
in \expressionname\eqref{equ:Ng_iter_seed_2} is crucial for numerical stability at strong particle correlations,
generalizing to quadratic and higher orders of polynomial extrapolation for $n_1>2$ seems to have a weakening effect on numerical stability. 

Variations of Ng's fixed point iteration method have been successfully applied in various liquid integral equation studies
\cite{Ng1974, Gonzalez-Mozuelos1998, Kovalenko1999, Mendez-Alcaraz2000, Gonzalez-Mozuelos2001, Chavez-Paez2003,
Castaneda-Priego2003, Anta2005, Contreras-Aburto2010, Sergiievskyi2012}.
General performance figures of the algorithm in \expressionsname\eqref{equ:Ng_scheme}-\eqref{equ:extrap_coeff2}
are difficult to formulate, as its efficiency depends on the number of particle species and
spatial dimensions, and, most importantly, on the pair-potential parameters. All individual HNC solutions presented in the present paper 
took few minutes or less to be computed on an inexpensive personal computer.

Note that the fixed point iteration scheme in \expressionsname\eqref{equ:Ng_scheme}-\eqref{equ:extrap_coeff2}
is merely one among a wide variety of solution methods that have been developed. Part of the alternative solution methods have
been reported to show superior numerical efficiency, at the cost of a more complicated implementation.
Along with Ng's algorithm, Newton-Raphson-like fixed-point iteration schemes,
first introduced by Gillan\cite{Gillan1979}, Lab\'{i}k \textit{et al.}\cite{Labik1985}, and Zerah\cite{Zerah1985}, are routinely used in integral equation studies
of liquids with strong pair-correlations \cite{Kahl1988, Belloni1988, Fushiki1988, Thalmann2000, Belloni2002, Carbajal-Tinoco2002, Fantoni2003,
Pastore2005, Sausset2009, Fantoni2009,  Brandt2010, Puibasset2012}.
For an elaborate comparison of Ng's and Zerah's methods, including a formulation of the latter method for liquid mixtures, we refer to 
appendix A of \refname{}\cite{Mendez-Alcaraz-Thesis}.
Furthermore, highly elaborate Newton-GMRES (Krylov subspace) algorithms have been applied\cite{Booth1999}, and have been
combined with multigrid techniques\cite{Kelley2004}. Yet another alternative approach is the vector extrapolation method\cite{Homeier1995}.
For the sake of simplicity, in the present study we do not employ the methods laid out
in \refsname\cite{Gillan1979, Labik1985, Zerah1985, Booth1999, Kelley2004, Homeier1995}.
In future studies however, use of such elaborate fixed point solution methods,
in combination with the Fourier transform method in \sectionname\ref{sec:sub:Log}, might give access to liquid integral equation solutions
of the PM for even larger ion-size and charge asymmetries than accomplished in the present work.

\subsection{Fourier transform on logarithmic grids}\label{sec:sub:Log}

In this subsection, we present our numerical method of choice to approximate the forward- and backward Fourier transforms in
\expressionsname\eqref{equ:Fourier_x_to_y} and \eqref{equ:Fourier_y_to_x}, for functions $f(x)$ and $\tilde{f}(y)$ that are sampled on finite computational grids.
The method employed here has been devised in essence by Talman\cite{Talman1978}, and constitutes a sophisticated version of
the so-called quasi-fast Hankel transform method of Siegman \cite{Siegman1977}. It is based on the use of logarithmic variables,
\textit{i.e.}, computational grids of the form
\begin{equation}\label{equ:logarithmic_grids}
x_n = x_0 \exp\{nL/N\},\qquad
y_n = y_0 \exp\{nL/N\},
\end{equation}
with grid index $n$ in the range $-\lfloor{N/2}\rfloor \leq n \leq \lfloor{N/2}\rfloor$, and $L,N > 0$.
The use of logarithmic grids has been motivated by the work of Gardner \textit{et al.}\cite{Gardner1959}.
As demonstrated in \refname{}\cite{Siegman1977}, sampling on logarithmic grids allows to re-write \expressionsname\eqref{equ:Fourier_x_to_y}
and \eqref{equ:Fourier_y_to_x} as discrete circular correlations, each of which can be treated by applying two subsequent fast Fourier transforms (FFTs). 
For a complete, and particularly clear-cut documentation, we refer to the work of Hamilton\cite{Hamilton2000, Hamilton_website},
where the method has been named FFTLog. Since we have closely followed \refsname\cite{Hamilton2000, Hamilton_website} in our implementation of the FFTLog transform,
we refrain from repeating all details here, and list only the essential expressions instead.

Adapting to the notation of \refsname\cite{Hamilton2000, Hamilton_website}, we 
define the primed sum symbol $\sum'$ through
\begin{equation}\label{equ:primed_sum_symbol}
{\sum\limits_n}' x_n = \sum\limits_{n=-\lfloor{N/2}\rfloor}^{\lfloor{N/2}\rfloor} w_n x_n, 
\end{equation}
with weights $w_n = 1$ for all $n$, except for $w_{-N/2} = w_{N/2} = 1/2$ if $N$ is even.

In a preprocess, preceding the many FFTLog transforms occurring in the fixed point iteration described in the previous section,
lookup tables of the grid-specific coefficients
\begin{equation}\label{equ:FFTLog_coeffs}
u_n = {\left(\frac{2}{x_0 y_0} \right)}^{2\pi i n / L}
\Gamma\left[\frac{d}{4}+\frac{\pi i n}{L}\right] /
\Gamma\left[\frac{d}{4}-\frac{\pi i n}{L}\right]
\end{equation}
are computed. As pointed out in \refsname\cite{Talman1978, Caillol1981}, numerical evaluation of \expressionname\eqref{equ:FFTLog_coeffs} for large 
arguments of the Gamma functions is considerably simplified by noticing that $| \Gamma(x+iy)/\Gamma(x-iy) | = 1$ for $x,y \in \mathbb{R}$. 
The remaining problem of determining the complex phases of the Gamma functions in \expressionname\eqref{equ:FFTLog_coeffs}
is conveniently solved in resorting to \refname{}\cite{GSL}. An alternative way to determine the complex phases has been devised in \refname\cite{Caillol1981},  
where the problem was tackled using recurrence relations of the Gamma function.

The forward FFTLog transform from the space of dimensionless distances $x$ to the space of
dimensionless wavenumbers $y$ can be evaluated as \cite{Hamilton2000, Hamilton_website}
\begin{subequations}\label{equ:FFT_forward}
\begin{eqnarray}
\tilde{f}(y_n) &=& {\left({\frac{2\pi}{y_n}}\right)}^{d/2}{\sum\limits_m}' c_m u_m \exp\left\lbrace\frac{-2\pi i mn}{N}\right\rbrace,\label{equ:FFT_forward_1}\\
c_m &=& \frac{1}{N}{\sum\limits_n}'f(x_n) x_n^{d/2} \exp\left\lbrace\frac{-2\pi i mn}{N}\right\rbrace,~~~~\label{equ:FFT_forward_2} 
\end{eqnarray}
\end{subequations}
and the inverse transform (from y- to x-space) reads \cite{Hamilton2000, Hamilton_website}
\begin{subequations}\label{equ:FFT_inverse}
\begin{eqnarray}\label{equ:FFT_backward}
{f}(x_n) &=& x_n^{-d/2} {\sum\limits_m}' \frac{\tilde{c}_m}{u_m^*} \exp\left\lbrace\frac{-2\pi i mn}{N}\right\rbrace,\label{equ:FFT_inverse_1}\\
\tilde{c}_m &=& \frac{1}{N}{\sum\limits_n}'\tilde{f}(y_n) {\left({\frac{\displaystyle{y_n}}{\displaystyle{2\pi}}}\right)}^{d/2}
\exp\left\lbrace\frac{-2\pi i mn}{N}\right\rbrace,~~~~\label{equ:FFT_inverse_2} 
\end{eqnarray}
\end{subequations}
with the star denoting the complex conjugate.
Being nothing else  but discrete, one-dimensional Fourier transforms, all four \expressionsname\eqref{equ:FFT_forward_1}-\eqref{equ:FFT_inverse_2}
can be solved by numerically efficient FFT algorithms, which are available in a great variety of implementations. Here, we employ a mixed-radix FFT routine
\cite{GSL}, imposing no constraints on the number of grid points $2\lfloor N/2 \rfloor + 1$.

For validity of \expressionsname\eqref{equ:FFT_forward_2} and \eqref{equ:FFT_inverse_2},
it is necessary to fulfill the constraint $u_{-\lfloor N/2 \rfloor} = u_{\lfloor N/2 \rfloor}$
in choosing the parameters $x_0, y_0, N$, and $L$ of the grids in \expressionname\eqref{equ:logarithmic_grids}.
Fulfilling this constraint is equivalent to
\begin{equation}\label{equ:low_ringing}
\frac{N \ln(x_0 y_0)}{L} = \frac{1}{\pi} \text{Arg}\left[ 2^{\pi i N/L}
\frac{\displaystyle{\Gamma(d/4 + \pi i N/(2L))}}{\displaystyle{\Gamma(d/4 - \pi i N/(2L))}}
\right] + z, 
\end{equation}
with an arbitrary integer number $z$, and \mbox{Arg$[c]$} denoting the phase of complex number $c$.
In \refsname\cite{Hamilton2000, Hamilton_website}, the criterion in \expressionname\eqref{equ:low_ringing}
has been named the low-ringing condition. In all calculations with results presented here, we have chosen $x_0=1$,
and $x_0y_0 \approx 1$, while fulfilling \expressionname\eqref{equ:low_ringing}.

An important virtue of the FFTLog transform in \expressionsname\eqref{equ:FFTLog_coeffs}-\eqref{equ:FFT_inverse_2} is its
computational efficiency for arbitrary dimensions $d$. Choosing the number of grid points as an integer power of $2$
results in optimal performance of the transforms in \expressionsname\eqref{equ:FFT_forward_1}-\eqref{equ:FFT_inverse_2},
each requiring $\mathcal{O}(N\log_2 N)$ arithmetic operations in that case.
We note here that an alternative, numerically efficient method of calculating $d$-dimensional Fourier-Bessel transforms of the kind of
\expressionsname\eqref{equ:Fourier_x_to_y} and \eqref{equ:Fourier_y_to_x} has been used in
\refsname\cite{Blum1973, Patey1978, Fries1985, Klapp2000}. In this alternative approach, which does not
require logarithmic grids, the Fourier-Bessel transforms are replaced by a sequence of so-called hat transforms and FFTs.

For even $d$, in particular for $d=2$, numerically less efficient methods for computing the transforms in \expressionsname\eqref{equ:Fourier_x_to_y}
and \eqref{equ:Fourier_y_to_x} have been reported\cite{Lado1968, Lado1971, Guizar-Sicairos2004}, each of which requires
$\mathcal{O}(N^2)$ arithmetic operations. Such numerically sub-optimal $\mathcal{O}(N^2)$ transforms have been applied in various
liquid integral equation studies\cite{Hoffmann2007, Camargo2008, Hajnal2011}.

For $d=3$ (odd number of dimensions, in general), using the transform in \expressionsname\eqref{equ:primed_sum_symbol}-\eqref{equ:low_ringing}
is not obvious since the standard FFT can be directly applied to functions sampled on grids with uniform spacing.
Uniformly spaced grids, however, are not ideally suited for sampling
the correlation functions of highly asymmetric PM fluids. Ion size- and charge-ratios of the order of 1:1000 in typical colloidal suspensions
render it necessary to simultaneously resolve length scales that differ by a factor of more than one thousand, which requires huge numbers of grid points
in uniformly spaced grids.
For instance, in \refname{}\cite{Leger2005}, $2^{18} = 262 144$ points had to be used to sample correlation functions for ion diameter asymmetry of 1:80 and
charge asymmetries up to 1:450 (at non-zero macroion number density), resembling only rather small macroions.
Logarithmic grids, on the other hand, are ideally suited to capture the different length scales in (asymmetric) charged sphere systems,
as it has been first pointed out by Rossky and Friedman\cite{Rossky1980}. 
Logarithmic grids have later been used in $d=3$ liquid integral equation studies \cite{Kalyuzhnyi2001, Lin2002, Kloss2008} but,
to our knowledge, no liquid integral equation study
has been conducted so far, where the advantages of the the transform in \expressionsname\eqref{equ:primed_sum_symbol}-\eqref{equ:low_ringing}
have been exploited in solving pair correlations of an extremely asymmetric PM.
Our results presented here, for ion charge- and size-ratios both simultaneously as high as 1:1000, have been obtained
using no more than $2^{13} = 8192$ grid points.

\subsection{Molecular Dynamics Simulations}\label{sec:sub:MD}

In this paper, we present MD simulation data for fluids in $d=3$ spatial dimensions only.
We have simulated globally electroneutral systems with three or four different ion species
in a cubic box of edge length $B$ with periodic boundary conditions in all three Cartesian directions. 
The MD simulation method used here is the same as in \refsname{}\cite{Allahyarov1998, Allahyarov2009}.
In order to handle the long-ranged Coulomb interactions, the Lekner summation method \cite{Lekner1989, Lekner1991,Mazars2001} is employed.

\begin{figure}
\centering
\vspace{-4em}
\includegraphics[width=1\columnwidth]{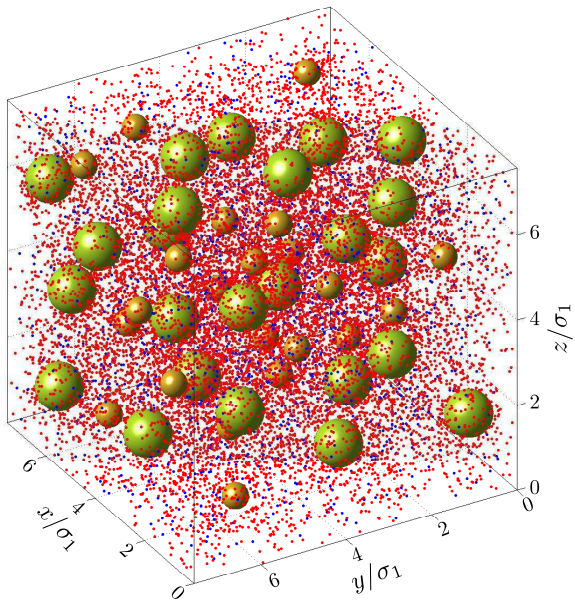}
\vspace{-3em}
\caption{\label{fig:Snapshot}
A representative snapshot of particle positions from our MD simulations of a four-component, three dimensional PM
(right column of \tablename{}~\ref{tab:simparam}, with rdf's shown in \figurename{}\ref{fig:HNC_AllahyarovMC_Quaternary_PM}).
The full simulation box is shown. Two macroion species (green and bronze) of different diameters and charge numbers are contained.
Sizes of coions (blue) and counterions (red) are exaggerated, to render the microions visible. 
}
\end{figure}

All parameters for the simulations presented here are listed in \tablename{}~\ref{tab:simparam},
with $N_i$ denoting the total number of particles of species $i$ in the simulation box, such that $n_i = N_i / B^3$.
All simulations are for a Bjerrum length of $L_B = 0.701$ nm, corresponding to water at room temperature.
A representative snapshot of particle positions for the four-component system (rightmost column of
\tablename{}~\ref{tab:simparam}) is shown in \figurename{}\ref{fig:Snapshot}.

\begin{table}
\caption{Parameters for the $d=3$ MD simulations of the present study. The $N_i$ are the numbers of
particles of species $i$ in the cubic box of edge length $B$.}
\vspace{.2em}
\centering
{
\begin{footnotesize}
\begin{tabular}{@{\extracolsep{\fill}}lll}
\hline
~~&\textbf{Ternary}\qquad\qquad\qquad&\textbf{Quaternary}\\ 
\hline\hline
\textbf{Results in}\qquad\qquad\qquad& \figurename{}\ref{fig:HNC_AllahyarovMC_Ternary_PM} & \figurename\ref{fig:HNC_AllahyarovMC_Quaternary_PM}\\
\hline
$\boldsymbol{N_1}$ & 48   & 24\\
$\boldsymbol{N_2}$ & 4080 & 24\\
$\boldsymbol{N_3}$ & 6280 - 28080 & 1900\\
$\boldsymbol{N_4}$ & -.- & 13680\\
\hline
$\boldsymbol{\sigma_1}$ & 150 nm & 122 nm\\
$\boldsymbol{\sigma_2}$ & 0.60 nm & 68 nm\\
$\boldsymbol{\sigma_3}$ & 0.60 nm & 0.61 nm\\
$\boldsymbol{\sigma_4}$ & -.-     & 0.61 nm\\
\hline
$\boldsymbol{Z_1}$ & 25 - 500 & 380\\
$\boldsymbol{Z_2}$ & 1        & 190\\
$\boldsymbol{Z_3}$ & -1       & 1\\
$\boldsymbol{Z_4}$ & -.-      & -1\\
\hline
$\boldsymbol{B/\sigma_1}$ & 7.951 & 7.566\\
\hline
\end{tabular}
\end{footnotesize}
%}
}
\label{tab:simparam}
\end{table}

On average, one week of execution time on a 64-bit computer cluster is enough to get fairly 
good statistics for the macroion-macroion and the macroion-microion correlation functions,
for the systems consisting of up to 32,208 charged particles. Achieving comparably good  
microion-microion statistics would require even much longer execution times. Therefore, we restrict ourselves
here to comparisons of macroion-macroion and macroion-microion pair correlations obtained from MD simulations
and the HNC scheme. 
%%%%%%%%%%%%%%%%%%%%%%%%%%%%%%%%%%%%%%%%%%%%%%%%%%%%%%%%%%%%%%%%%%%%%%%%%%%%%%%%%%%%%%%%%%%%%%%%%%%%%%%%%%%%%%%%

\section{Results}\label{sec:Results}

\subsection{Thermodynamical properties and pair-correlations in one to six dimensions}\label{sec:sub:Dim1to6}

As a first result, the HNC solutions computed in our numerical algorithm
are in agreement with the $d$-dimensional local electroneutrality (LEN) conditions
\begin{equation}\label{equ:local_en}
\text{sgn}(Z_i) \sqrt{|\Gamma_{ii}|} \stackrel{!}{=} - \lim\limits_{x \to \infty} A(d) \int\limits_{0}^{x} dx' \tau_i(x') {x'}^{d-1},  
\end{equation}
where $\tau_i(x)$, defined as 
\begin{equation}\label{equ:local_chargedens}
\tau_i(x) = \chi_i \text{sgn}(Z_i Z_j) \sqrt{|\Gamma_{ij}|} g_{ij}(x)
\end{equation}
is the isotropic charge density around a test particle of species $i$.
In \expressionsname{}\eqref{equ:local_en} and \eqref{equ:local_chargedens}, $\text{sgn}(x)$ is the sign function.
The LEN condition states that the total charge of ions around a test particle cancels out with the test particle's charge.
If a computational grid is chosen that extends to a very large
outer radius $x_{\lfloor N/2 \rfloor}$, we find the LEN condition in \expressionname\eqref{equ:local_en} violated at large $x$, where
the functions $c_{ij}^{(s)}(x)$ and $\gamma_{ij}^{(s)}(x)$ assume values too small to be resolved at machine precision. However, if taking the
$x \to \infty$ limit in \expressionname\eqref{equ:local_en} is replaced by insertion of an intermediately large value of $x \approx 100$,
where the oscillations in all $g_{ij}(x)$ have essentially died out, we find \expressionsname\eqref{equ:local_en} fulfilled
to within good accuracy.

\begin{figure}
\centering
\includegraphics[width=.75\columnwidth,angle=-90]{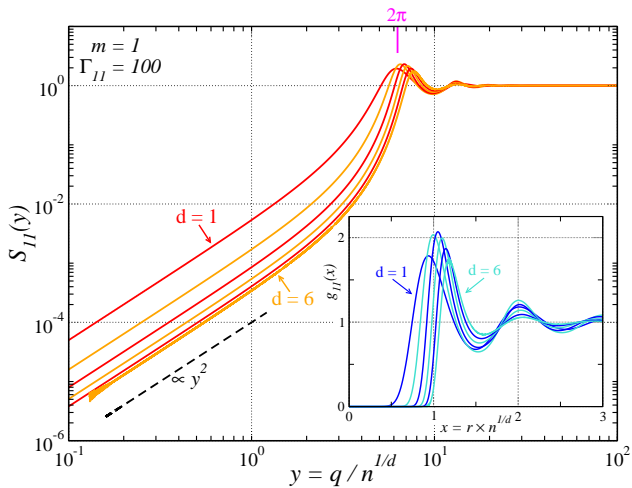}
\vspace{0em}
\caption{\label{fig:OCP_nd}
One component plasma ($m=1$) static structure factors, $S_{11}(y)$, and rdf's, $g_{11}(x)$, for coupling parameter $\Gamma_{11}=100$.
In each case, the hypervolume fraction $\phi_1$ has been chosen small enough to ensure a
practically vanishing rdf contact value, $g_{11}(x=\sigma_{11}n^{1/d}) \approx 0$.
Correlation functions for systems in all integer dimensions from $d=1$ to $d=6$ are plotted. Principal peak positions
of $S_{11}(y)$ and $g_{11}(x)$ shift from left to right as $d$ increases.
}
\end{figure}

In the special case of $m=1$, our algorithm allows to compute pair-correlations of the OCP in arbitrary dimensions, as 
illustrated in \figurename{}\ref{fig:OCP_nd}. Note here the good quality of the HNC solutions at small wavenumbers,
where the vanishing compressibility of the OCP, with $\lim_{y\to 0} S_{11}(y) = 0$, is well described.
The magnitude of the undulations in the $S_{11}(y)$ and $g_{11}(x)$ plotted in \figurename{}\ref{fig:OCP_nd} is a non-monotonic function of the dimension $d$,
with maximal undulations occurring for $d=3$.

\begin{figure*}
\centering
\includegraphics[width=.55\textwidth,angle=-90]{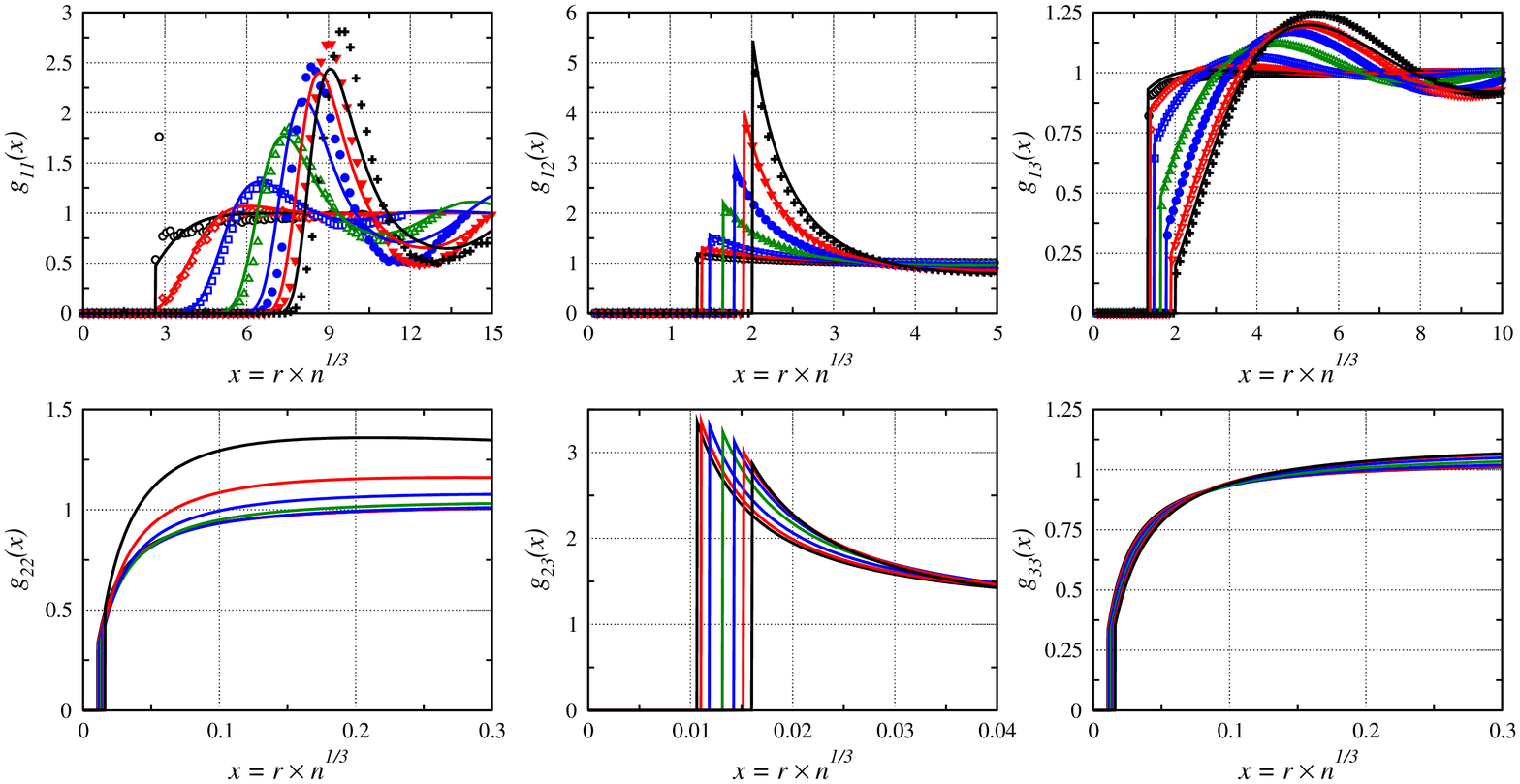}
\vspace{-1em}
\caption{\label{fig:HNC_AllahyarovMC_Ternary_PM}
HNC (solid curves) results for all partial rdf's of three-dimensional, three-component, globally electroneutral primitive models.
Macroion-macroion and macroion-microion rdf's are compared with MD simulation results (symbols) in the upper row of panels. Each panel corresponds
to a fixed particle-species pairing, as indicated on the vertical axes labels.
Common system parameters are
$\sigma_1 = 150$ nm, $\sigma_2 = \sigma_3 = \sigma_1 / 250 = 0.6$ nm, $L_B = 0.701$ nm, $\phi_1 = 0.05$, $n_3 = 4 \mu M$, $Z_2 = -1$, $Z_3 = 1$.
The macroion charge number, $Z_1$, has been varied, assuming the values
$Z_1 = 25$ (open black circles, black lines),
$Z_1 = 50$ (open red diamonds, red lines),
$Z_1 = 100$ (open blue squares, blue lines),
$Z_1 = 200$ (open green upwards triangles, green lines),
$Z_1 = 300$ (filled blue circles, blue lines),
$Z_1 = 400$ (filled red downward triangles, red lines), and
$Z_1 = 500$ (black crosses, black lines).
}
\end{figure*}

\begin{figure*}
\centering
% Twocolumn:
\includegraphics[width=.95\columnwidth,angle=-90]{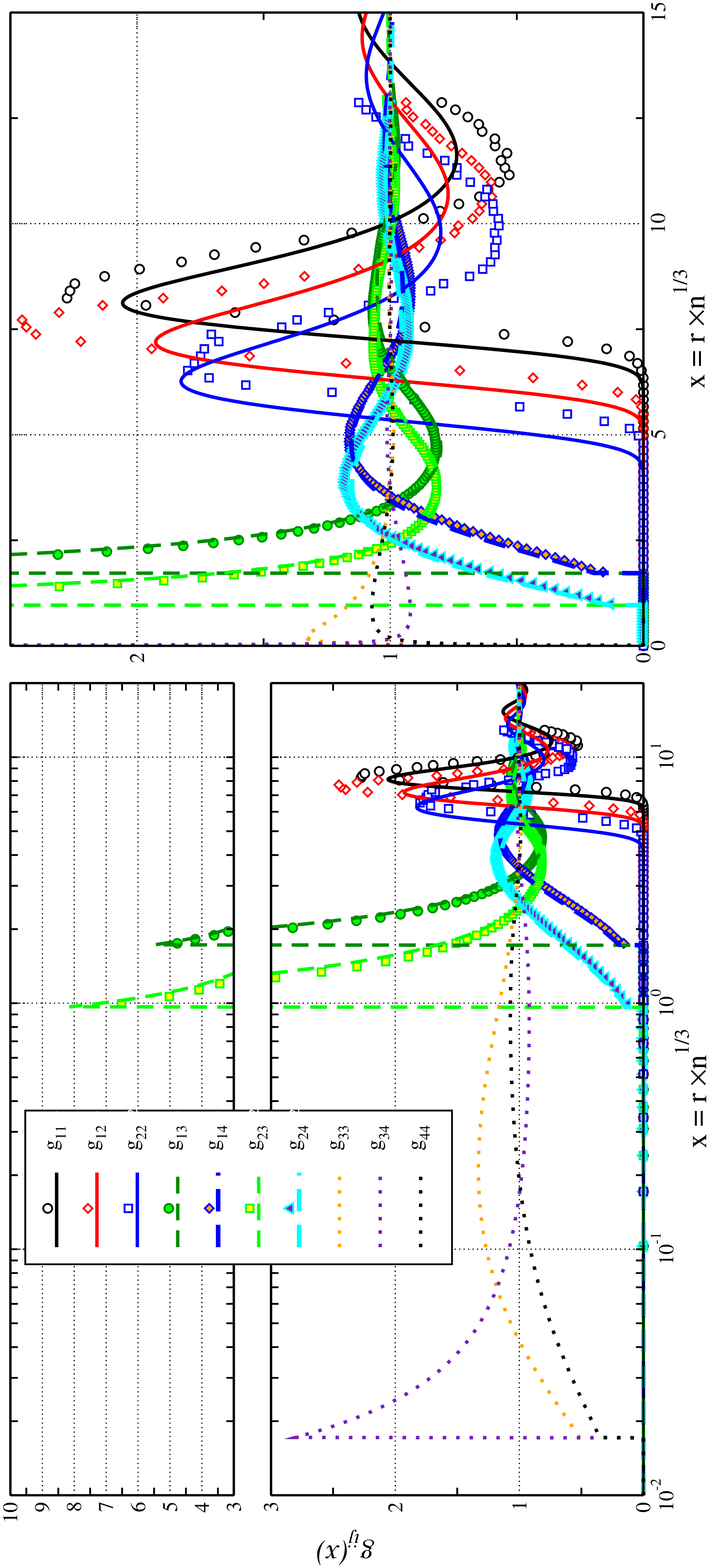}
% Preprint:
%\includegraphics[width=.45\columnwidth,angle=-90]{Figure4.eps}
%\vspace{-1em}
\caption{\label{fig:HNC_AllahyarovMC_Quaternary_PM}
HNC (solid, dashed, and dotted curves) results for all partial rdf's of a three-dimensional, four-component, globally electroneutral PM.
Macroion-macroion and macroion-microion rdf's are compared with MD simulation results (symbols).
Species $1$ and $2$ are macroions with charges of equal sign, where species $1$ is more strongly charged and possesses a larger
hard-core diameter than species $2$. Species $3$ are the counterions, which are of equal size, but opposite charge, as the coions of species $4$. 
System parameters are $\sigma_1 = 122$ nm, $\sigma_2 = 68$ nm, $\sigma_3 = \sigma_4 = \sigma_1 / 200 = 0.61$ nm,
$L_B = 0.701$ nm, $\phi = \sum_i \phi_i = 0.034$, $n_1 = n_2$, $n_4 = 4 \mu M$, $Z_1 = 380$ $Z_2 = 190$, $Z_3 = -1$, $Z_4 = 1$. 
The left two panels are in logarithmic-linear scale, and the right panel exposes the details of the macroion-macroion rdf's on a doubly linear scale.
}
\end{figure*}

In \figuresname{}\ref{fig:HNC_AllahyarovMC_Ternary_PM} and \ref{fig:HNC_AllahyarovMC_Quaternary_PM}, we compare the HNC rdf's $g_{ij}(r)$
for three- and four-component primitive models in $d=3$ to the results of our MD simulations.
Overall good agreement is observed between the HNC and MD results, the most prominent discrepancy being
an underestimation (of up to about $20\%$) of the principal peak height in the HNC macroion-macroion rdf's, occurring at strong macroion correlations.
Underestimation of the principal peak heights in the macroion-macroion pair-correlation functions is a known shortcoming of the HNC
\cite{Rogers1984, Hansen_McDonald1986, Heinen2011}, which can be tackled by choosing an alternative, thermodynamically
partially self-consistent integral equation scheme \cite{Zerah1986, Caccamo1996, Nagele1996, Caccamo1999}.
An alternative method to improve the accuracy of the HNC consists in using a tailored Ansatz \cite{Ng1974, Hajnal2011} for the bridge
function \cite{Hansen_McDonald1986} at high coupling, which is neglected altogether in the HNC.  

\begin{figure*}
\centering
\vspace{-0.5em}
\includegraphics[width=.95\textwidth]{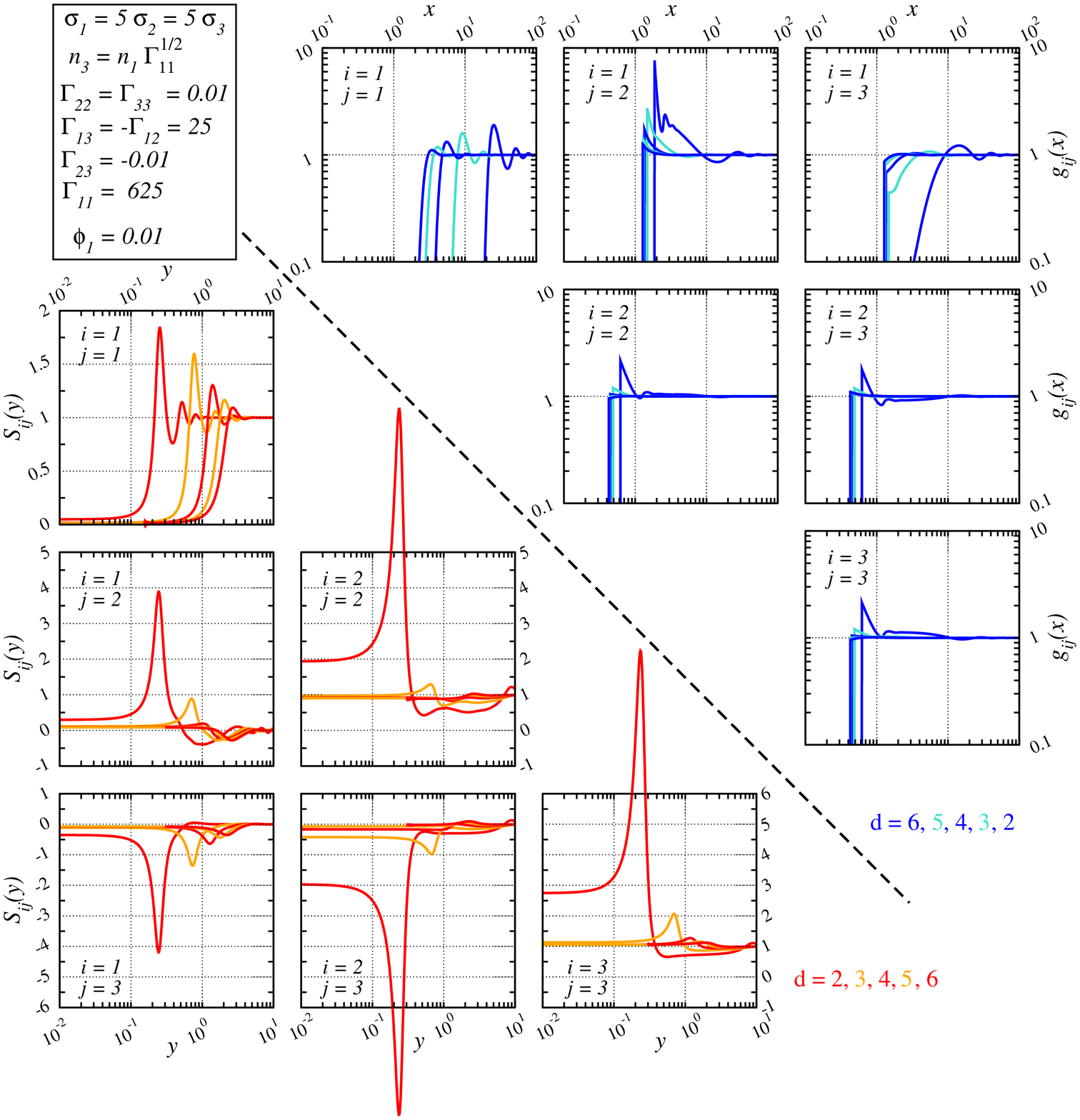}
\vspace{-1.5em}
\caption{\label{fig:gr_Sq_nd_PM}
HNC partial rdf's $g_{ij}(x)$ (dark and light blue) and partial static structure factors $S_{ij}(y)$ (red and orange)
of three-component, globally electroneutral primitive models in $d = 2, 3, 4, 5,$ and $6$
spatial dimensions, with equal dimensionless pair-potential parameters as indicated in the box in the upper left corner. The undulations in the functions are strongest
for $d=2$, and decay with rising dimension $d$. All rdf's are plotted in double-logarithmic scales of equal horizontal and vertical axes ranges.
The horizontal $y$-axis ranges are equal in all structure factor plots, and the vertical $S_{ij}(y)$-axis ranges are $[0,2]$ for $i=j=1$,
$[-1,6]$ for $j=2$ and for $i=j=3$, and $[-6,1]$ for $j=3\neq i$. The peak positions of the three $d=2$ partial static structure factors that exceed their panel's
vertical axis ranges are $S_{22}(y=0.237) = 9.0$, $S_{23}(y=0.237) = -9.5$, and $S_{33}(y=0.236) = 10.7$.}
\end{figure*}

In \figurename{}\ref{fig:gr_Sq_nd_PM} we display all HNC partial rdf's and static structure factors of three-component,
globally electroneutral primitive models in all integer dimensions $d$ from $2$ to $6$. The dimensionless pair-potential parameters are the same for
all systems in the figure, with coupling constants $\Gamma_{22} = \Gamma_{33} = -\Gamma_{23} = 0.01$, $\Gamma_{13} = -\Gamma_{12} = 25$,
and $\Gamma_{11} = 625$, corresponding, for $d=3$, to $Z_1 = 250$, $Z_2 = -1$, $Z_3 = 1$, and $L_B n^{1/3} = 0.1$. The hypervolume fraction of species $1$ is 
$\phi_1 = 0.01$ for all systems in \figurename{}\ref{fig:gr_Sq_nd_PM}, and $n_3 = n_1 \sqrt{\Gamma_{11}}$, which means that there is one
salt coion per macroion-surface released counterion. A rather small ratio of macroion- to microion diameters, $\sigma_1 / \sigma_2 = \sigma_1 / \sigma_3 = 5$,
has been chosen for the systems in \figurename{}\ref{fig:gr_Sq_nd_PM} since, for larger size asymmetry, the solution in higher dimensions such as $d=6$
becomes numerically very slowly convergent or divergent. Note that \figurename{}\ref{fig:gr_Sq_nd_PM} features panels with double logarithmic
as well as logarithmic-linear axes, and that the axes ranges vary from panel to panel, to exhibit simultaneously the details of the various plotted functions. 

In \figurename{}\ref{fig:gr_Sq_nd_PM}, the $g_{ij}(x)$ and $S_{ij}(y)$ with the most pronounced oscillations around the asymptotic value one are
for $d=2$. For rising dimension and fixed dimensionless potential parameters, particle packing becomes less efficient and the decay of the Coulomb potentials
becomes steeper. Therefore, the undulations in the pair-correlation functions get reduced for rising $d$ until, for $d=6$, the undulations have almost
completely died out.

\subsection{Application to the colloidal domain}\label{sec:sub:Coll}

\begin{figure}
\centering
% Twocolumn:
\includegraphics[width=.9\columnwidth]{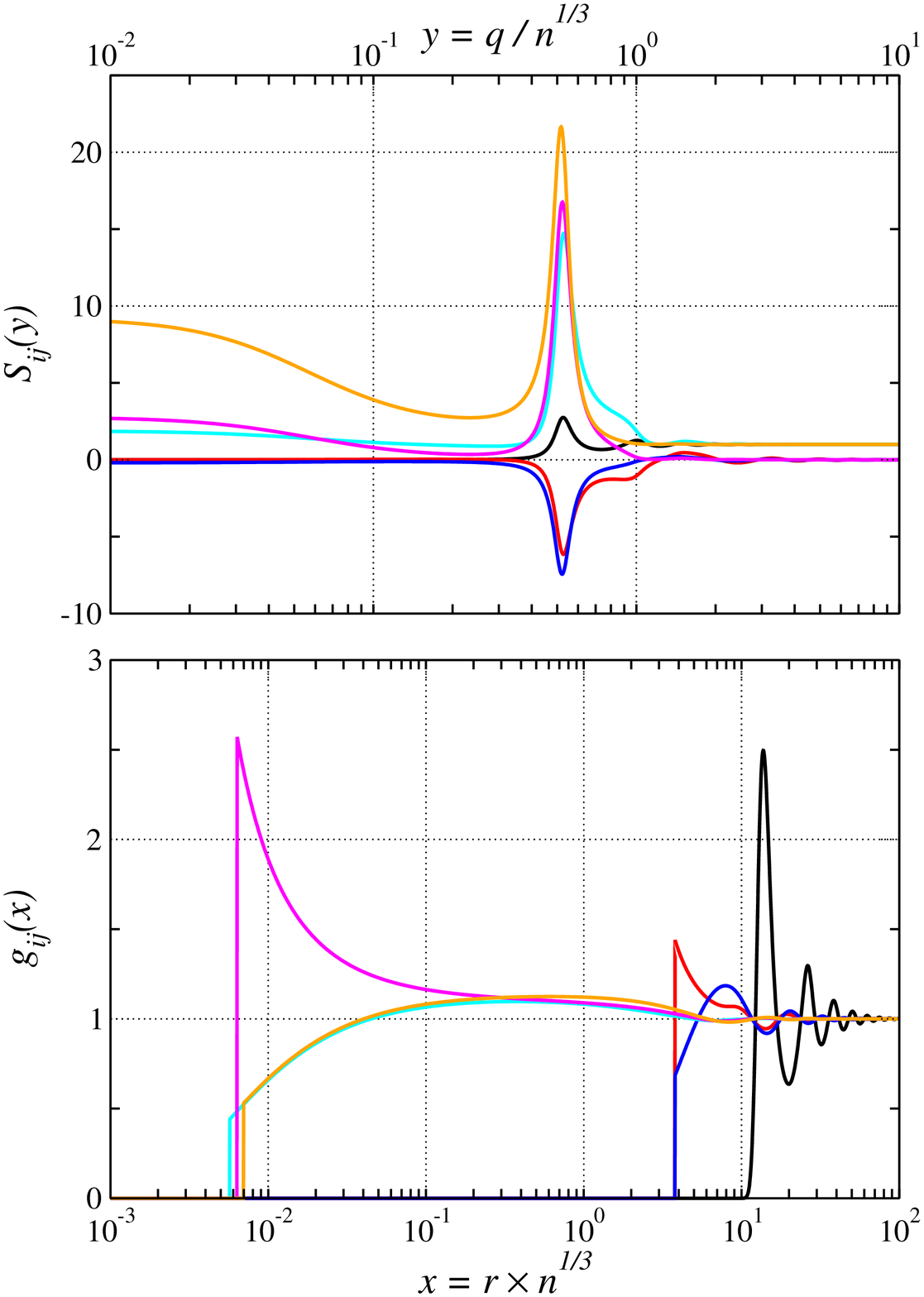}
% Preprint:
%\includegraphics[width=.65\columnwidth]{Figure6.eps}
%\vspace{-1em}
\caption{\label{fig:HNC_colloidal_micronsize_Ternary_PM}
All HNC partial static structure factors (top) and rdf's (bottom) for a three-dimensional, three-component primitive model of micrometer-sized colloids in
aqueous $NaCl$ electrolyte.
Parameters:
$\sigma_1 = 1000$ nm,
$\sigma_2 = 0.756$ nm (hydrated $Na+$), 
$\sigma_3 = 0.922$ nm (hydrated $Cl-$), 
$Z_1 = -750$,
$Z_2 = 1$,
$Z_3 = -1$,
$\phi_1 = 0.1$ ,
$L_B = 0.70$ nm,
$n_3 = n_1 |Z_1|$ (one coion per surface-released counterion).
}
\end{figure}

Figure \ref{fig:HNC_colloidal_micronsize_Ternary_PM} features all HNC $g_{ij}(x)$ and $S_{ij}(y)$ for a $d=3$, ternary PM that resembles
a realistic suspension of colloidal particles in aqueous ($L_B = 0.70$ nm) electrolyte with a low concentration of dissociated $NaCl$. The diameter of the macroions
is taken to be $\sigma_1 = 1 \mu m$, corresponding to rather large colloidal particles, while the diameters $\sigma_2 = 0.756$ nm and 
$\sigma_3 = 0.922$ nm correspond to hydrated $Na+$ and $Cl-$ ions, respectively.
Here, we have chosen the concentration of $Cl-$ coions (species 3) as $n_3 = n_1 |Z_1|$, so that the suspension contains
one coion per colloid-surface released counterion, and overall about twice as many counterions as coions, $n_2 \approx 2 n_3$.
With the assumed macroion charge $Z_1 = -750$, which is a realistic bare charge for a micron-sized colloidal sphere,
this gives a coion concentration of $n_3 = 0.71 \mu M$. This corresponds to an almost deionized aqueous solvent with
a little amount of dissociated salt only (\textit{c.f.}, the number concentration, $n = 0.1 \mu M$,
of the water self-dissociation products $H_3 O+$ and $OH-$ at neutral pH-value, which is a lower bound for the coion concentration).

\subsection{Effective colloidal interactions}\label{sec:sub:Eff_int}
%%%%%%%%%%%%%%%%%%%%%%%%%%%%%%%%%%%%%%%%%%%%%%%%%%%%%%%%%%%%%%%%%%%%%%%%%%%%%%%%%%%%%%%%%%%%%%%%%%%%%%%%%%%%%%%%

As a dimensionless effective pair potential between particles of the same species $a$ in an $m$-component fluid mixture, one can define \cite{Fushiki1988}
\begin{equation}\label{equ:inverted_HNC}
u_{aa}^{\text{eff}}(x) = h_{aa}(x) - c_{aa}^{\text{eff}}(x) - \ln\left[g_{aa}(x)\right],
\end{equation}
where $c_{aa}^{\text{eff}}(x)$ is an effective direct correlation function between particles of species $a$.
The Fourier transform of the latter is
\begin{equation}\label{equ:effective_direct_corr_func}
\tilde{c}_{aa}^{\text{eff}}(y) = \frac{\tilde{h}_{aa}(y)}{1 + \chi_a \tilde{h}_{aa}(y)},
\end{equation}
with the total correlation function $\tilde{h}_{aa}(y) = \tilde{\gamma}_{aa}(y) + \tilde{c}_{aa}(y) = \tilde{\gamma}^{(s)}_{aa}(y) + \tilde{c}_{aa}^{(s)}(y)$
taken from the solution of the coupled $m$-component set of \expressionsname\eqref{equ:OZ_short}-\eqref{equ:u_trans_long}.
Equations \eqref{equ:inverted_HNC} and \eqref{equ:effective_direct_corr_func} constitute an inversion of the HNC for species $a$ only,
meaning that a one-component fluid of particles with pair-potential $\beta u_{aa}^{\text{eff}}(x)$,
solved within the HNC approximation, shows exactly the same pair-correlation functions
as component $a$ of the $m$-component mixture.

Consider now a three-dimensional, ternary ionic liquid mixture of macroionic spheres (species $a$) with diameter $\sigma_a$ and
charge number $Z_a$, monovalent counterions, and monovalent coions.
%Each counterion carries one elementary charge of sign opposite to that of $Z_a$,
%and each coion carries one elementary charge of sign equal to that of $Z_a$.
In this case, the repulsive part of the dimensionless Derjaguin-Landau-Verwey-Overbeek (DLVO) effective
pair potential between two macroions at a non-overlap center-to-center distance $x > \sigma_a n^{1/3}$
can be written as \cite{Verwey_Overbeek1948}
\begin{equation}\label{equ:DLVO_potential}
\beta u_{aa}^{\text{DLVO}}(x) =
\Gamma_{aa}~
\frac{\displaystyle{e^{\displaystyle{k \sigma_a n^{1/3}}}}}{{\left( 1 + \frac{\displaystyle{k \sigma_a n^{1/3}}}{\displaystyle{2}\rule[0em]{0em}{.9em}}\right)}^2}~ 
\frac{\displaystyle{e^{-kx}}}{\displaystyle{x}},
\end{equation}
with $\Gamma_{aa} = L_B n^{1/3} Z_{a}^2$, as defined further up this text.
Equation \eqref{equ:DLVO_potential} involves the dimensionless DLVO screening parameter $k$, which is given by  
\begin{equation}\label{equ:DLVO_screening}
k^2 = 4\pi L_B n^{1/3} \left(\chi_a |Z_a| + 2 \chi_{\text{coion}} \right)
\end{equation}
%2
with $\chi_{\text{coion}}$ denoting the mole fraction of coions. 
The DLVO potential in \expressionname\eqref{equ:DLVO_potential} is valid for two macroions in a bath of microions whose
distribution can be treated in the Debye-H\"{u}ckel approximation. It is thus valid only at low macroion concentration, and
for $L_B Z_a / \sigma_a \lesssim 1$, \textit{i.e.}, for ionic pair interactions that do not considerably exceed the thermal energy.
Under conditions where $L_B Z_a / \sigma_a > 1$, the potential in \expressionname{}\eqref{equ:DLVO_potential} is nevertheless
a good approximation to the effective macroion pair-potential at sufficiently large particle separation,
provided that the charge number $Z_a$, entering via $\Gamma_{aa}$ and \expressionname{}\eqref{equ:DLVO_screening}, is replaced by an effective
charge number $Z_a^{\text{eff}} < Z_a$
\cite{Alexander1984, Levin1998, Tamashiro1998, Diehl2001, Bocquet2002, Trizac2002, Trizac2003, Trizac2004, Pianegonda2007, Torres2008, McPhie2008, Colla2009}.
The effective charge $Z_a^{\text{eff}} e$ has to be regarded as the net charge of a colloid dressed with closely associated counterions.

\begin{figure}
\centering
\includegraphics[width=.7\columnwidth,angle=-90]{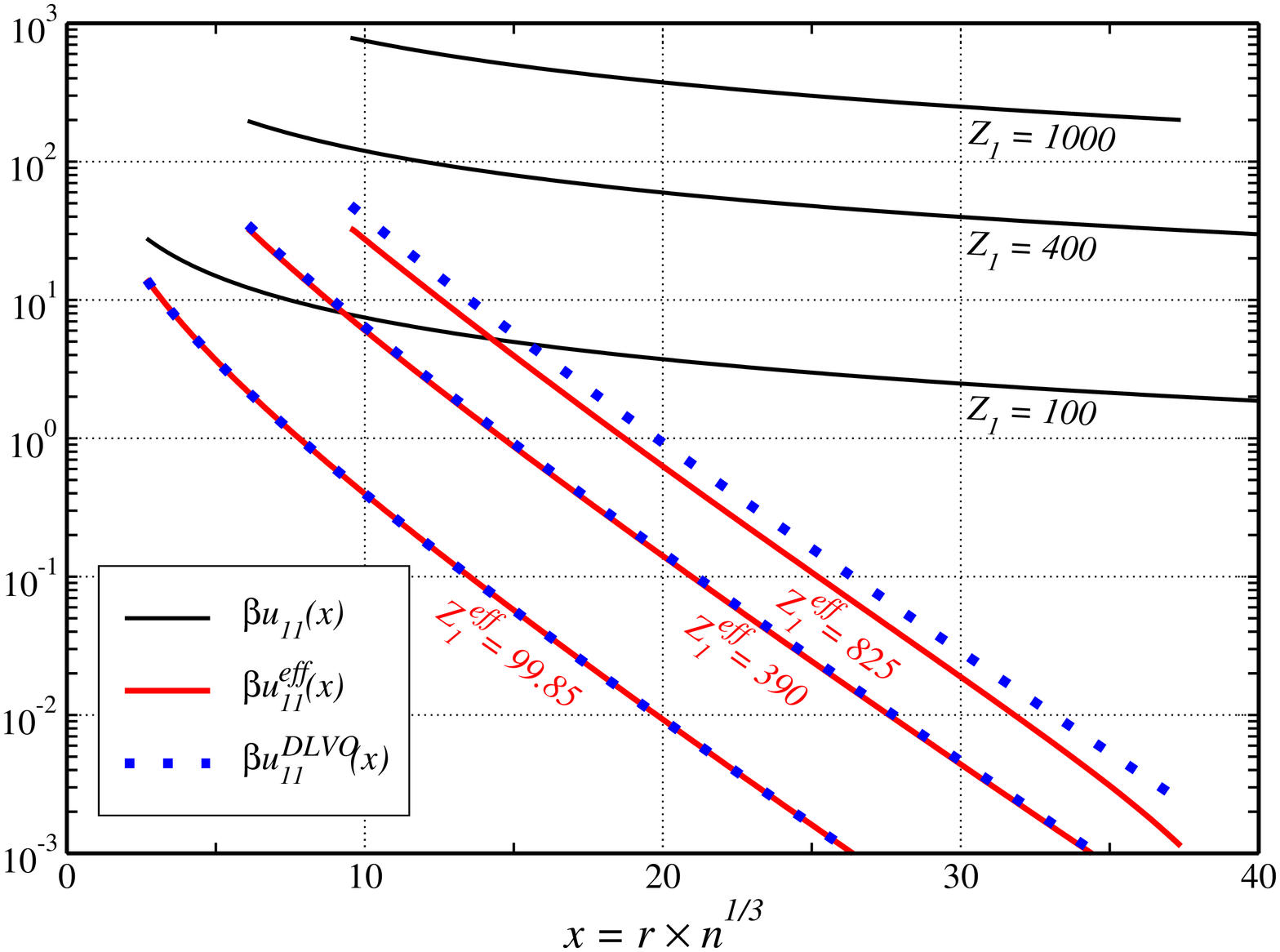}
%\vspace{-1em}
\caption{\label{fig:effpot}
Unscreened Coulomb-potentials $\beta u_{11}(x)$ (black solid curves),
effective pair-potentials $\beta u_{11}^{\text{eff}}(x)$,
computed by the HNC inversion in \expressionsname{}\eqref{equ:inverted_HNC} and \eqref{equ:effective_direct_corr_func} (red solid curves),
and DLVO pair potentials $\beta u_{11}^{\text{DLVO}}(x)$,
defined in \expressionsname\eqref{equ:DLVO_potential} and \eqref{equ:DLVO_screening} (blue dotted curves),
between a pair of macroions (species $1$) in three-component,
three-dimensional, globally electroneutral ionic mixtures with common parameters
$\sigma_1 = 250$ nm, $\sigma_2 = \sigma_3 = 0.6$ nm, $Z_2 = -1$, $Z_3 = 1$, $L_B = 0.701$ nm, $\phi_1 = 10^{-4}$, and $n_3 = 1\mu M$.
Results for three macroion charge numbers, $Z_1 = 100, 400$, and $1000$ are shown. 
}
\end{figure}
\begin{figure}
\centering
\includegraphics[width=.7\columnwidth,angle=-90]{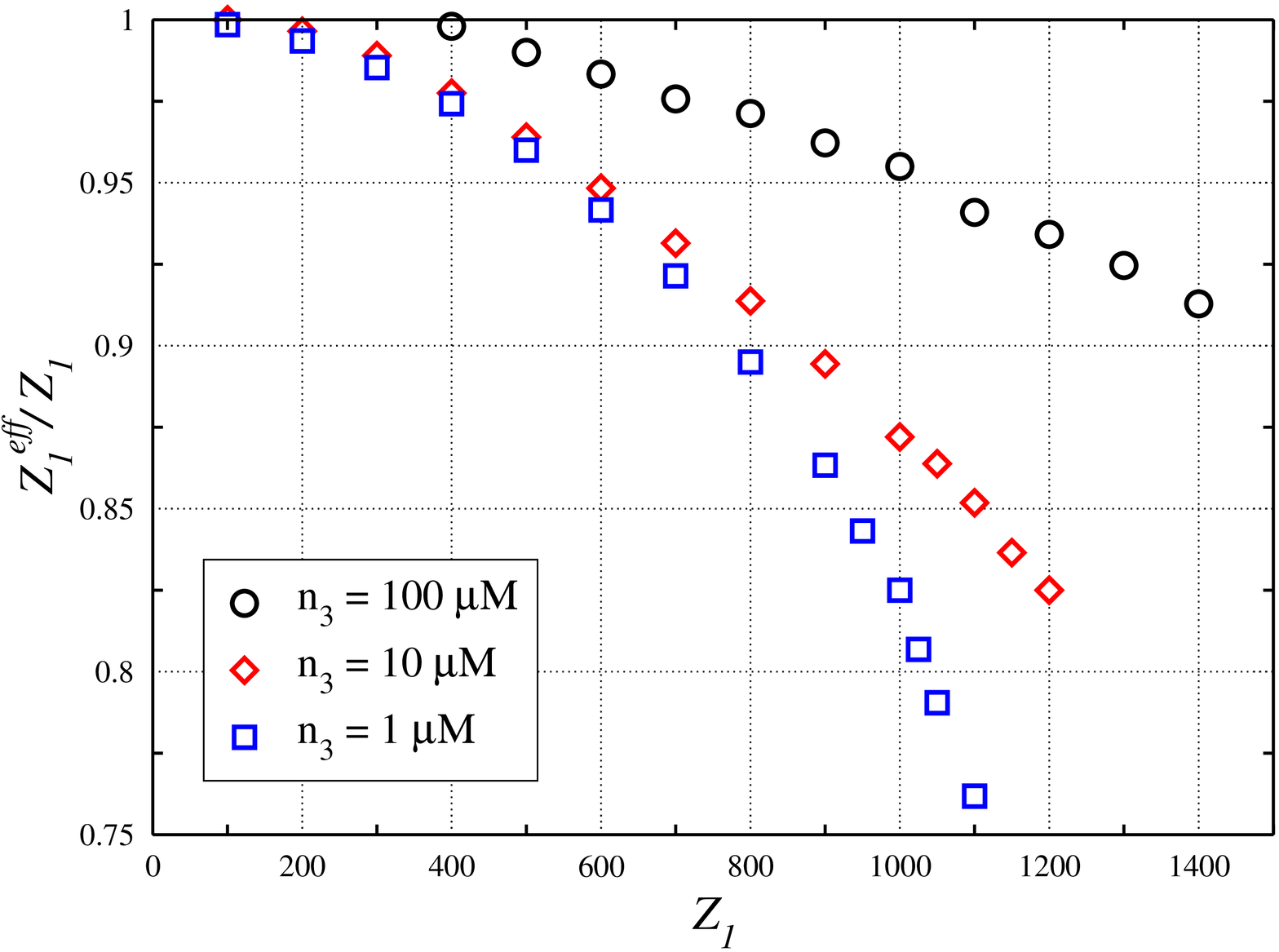}
%\vspace{-1em}
\caption{\label{fig:Zeff}
Reduced colloidal effective charge number, $Z_1^{\text{eff}}/Z_1$, as a function of the colloidal bare charge number $Z_1$, for   
three-dimensional, globally electroneutral ionic mixtures with
counterion concentrations $n_3 = 100\mu M$ (black circles), $n_3 = 10\mu M$ (red diamonds), and $n_3 = 1\mu M$ (blue squares). 
Common parameters are $\sigma_1 = 250$ nm, $\sigma_2 = \sigma_3 = 0.6$ nm, $Z_2 = -1$, $Z_3 = 1$, $L_B = 0.701$ nm, and $\phi_1 = 10^{-4}$.
}
\end{figure}

While theories for the saturation value, $Z_a^{\text{eff}}(Z_a \to \infty)$, of the effective charge number
are available \cite{Ohshima1982, Bocquet2002, Trizac2002},
calculation of the non-saturated effective charge in a PM, in its full dependence on the concentrations and charges of all ionic species,
remains a challenging task \cite{Alexander1984, Trizac2004, Allahyarov2005, Colla2009}.

We determine the effective charge by replacing $Z_a$ with $Z_a^{\text{eff}}$ in \expressionsname\eqref{equ:DLVO_potential} and
\eqref{equ:DLVO_screening}, and tuning $Z_a^{\text{eff}}$ until the resulting pair-potential optimally fits the HNC inversion
effective potential in \expressionname\eqref{equ:inverted_HNC} at large $x$. Examples for systems of colloidal particles with
$\sigma_a = \sigma_1 = 250$ nm, suspended at $\phi_a = \phi_1 = 10^{-4}$ in an aqueous 1-1 electrolyte with a fixed concentration,
$n_3 = 1\mu M$, of salt coions, are shown in \figurename{}\ref{fig:effpot}. Results for three different colloidal charge numbers, $Z_1 = 100, 400$, and
$1000$ are shown. Note that $\chi_3 = \chi_{\text{coion}} \gg Z_a \chi_a = Z_1 \chi_1$ for all three systems shown in
\figurename{}\ref{fig:effpot}, such that the DLVO screening
length in \expressionname\eqref{equ:DLVO_screening} is not considerably altered by replacing $Z_a$ with $Z_a^{\text{eff}}$. Fitting the potential
in \expressionname\eqref{equ:DLVO_potential} to the one obtained from \expressionname\eqref{equ:inverted_HNC} therefore corresponds to vertical
translation of $\beta u_{aa}^{\text{DLVO}}(x)$ in the linear-logarithmic plot of \figurename{}\ref{fig:effpot}. 
For the lowest considered colloid charge number, $Z_1=100$,
the DLVO-potential calculated according to \expressionsname\eqref{equ:DLVO_potential} and \eqref{equ:DLVO_screening} (lowermost blue dotted curve)
is in nearly perfect agreement with the HNC-inversion potential (lowermost red solid curve). Fitting the DLVO potential with effective
charge number input to the HNC-inversion potential, results in $Z_1^{\text{eff}} = 99.85$, which is only slightly lower than the bare
colloidal charge number. The same procedure, carried out for $Z_1 = 400$ and $Z_1 = 1000$, results in effective charge numbers
of $Z_1^{\text{eff}} = 390$ and $825$, respectively. This demonstrates the capability of the employed numerical methods to access HNC solutions
of the PM for realistic suspensions of charged colloids, where charge renormalization plays an important role.
In \figurename{}\ref{fig:Zeff}, we plot the effective charge of macroions (species 1) as a function of the macroion bare charge, for various
concentrations of salt coions, $n_3 = 1, 10,$ and $100 \mu M$. In agreement with a recent small angle x-ray scattering study for
charged silica spheres in aqueous electrolyte \cite{Westermeier2012}, we find that association of counterions to the macroion
surfaces is most efficient at low salinity. This is reflected in \figurename{}\ref{fig:Zeff} by a steepening decay of $Z_1^{\text{eff}}(Z_1)$, for increasing
salinity $n_3$.

\section{Conclusions}\label{sec:Conclusions}

We have shown that a combination of a numerically robust fixed-point iteration scheme and logarithmically spaced computational
grids allows efficient computation of HNC solutions of the $d$-dimensional primitive model, with explicit results shown for $d \leq 6$.
Logarithmic grids are ideally suited for the discretization of pair-correlation
functions of ionic mixtures with large particle diameter and charge asymmetries. This has allowed us to access HNC solutions
for primitive model parameters corresponding to realistic suspensions of micrometer-sized colloidal spheres with charge numbers as high as
$|Z| \approx 1000$, in an aqueous 1-1 electrolyte.

Numerical stability might be further improved in future studies, if another elaborate fixed-point iteration scheme is
used \cite{Gillan1979, Labik1985, Zerah1985, Booth1999, Kelley2004, Homeier1995}. We expect that this 
would give access to HNC solutions of the primitive model at charge and diameter asymmetries exceeding the ones reported here. 

Future projects, based on the methods presented here, might include \textit{ab initio} modeling of 
colloidal suspensions with reactive electrolytes such as $NaOH$ \cite{Herlach2010, Wette2010}, including chemical
association-dissociation reactions influenced by the locally varying pH-Value near the colloidal particle's surfaces. 

As opposed to molecular dynamics simulations of the asymmetric primitive model, with very long program execution times even for
moderate ion size- and charge-asymmetries, the solution of the HNC equations with the method described here takes
few minutes or less on an inexpensive, standard computer. In addition, the HNC is a good approximation for
mixtures of charged particles with long-ranged pair-potentials, predicting pair-correlation functions in good agreement with
the numerically expensive computer simulations. Despite continuing rapid progress in computer simulations, 
liquid integral equations therefore remain an indispensable approach in studying highly asymmetric electrolytes.

\subsection{Acknowledgements}

We gratefully acknowledge Martin Oettel and Thomas Palberg for helpful discussions. 
This work was funded by the ERC Advanced Grant INTERCOCOS, FP7 Ref.-Nr. 267499.

%\clearpage

%%%%%%%%%%%%%%%%%%%%%%%%%%%%%%%%%%%%%%%%%%%%%%%%%%%%%%%%%%%%%%%%%%%%%%%%%%%%%%%%%
% BIBLIOGRAPHY

\bibliographystyle{unsrt}
\bibliography{ARXIV_HNC_PM}   % Produces the bibliography via BibTeX.

%%%%%%%%%%%%%%%%%%%%%%%%%%%%%%%%%%%%%%%%%%%%%%%%%%%%%%%%%%%%%%%%%%%%%%%%%%%%%%%%%
%\clearpage
%%%%%%%%%%%%%%%%%%%%%%%%%%%%%%%%%%%%%%%%%%%%%%%%%%%%%%%%%%%%%%%%%%%%%%%%%%%%%%%%%

\end{document}